\RequirePackage{fix-cm}
\documentclass{svjour3}                     % onecolumn (standard format)
\smartqed  % flush right qed marks, e.g. at end of proof
\usepackage{graphicx}
\usepackage{natbib}

\usepackage{amsmath,amssymb}

\renewcommand{\d}{{\rm d}}

\newcommand{\Z}{{\bf Z}}

\newcommand{\R}{{\mathbb{R}}}
\newcommand{\e}{{\rm e}}
\newcommand{\sgn}{{\rm sgn}}
\newcommand{\D}{\displaystyle}
\newcommand{\Db}{{\bf D}}
\newcommand{\ve}{\varepsilon}
\newcommand{\mc}{\mathcal}
\newcommand{\bxi}{{\boldsymbol \xi}}
\newcommand{\bthet}{{\boldsymbol \theta}}
\newcommand{\bvthet}{{\boldsymbol \vartheta}}
\newcommand{\ab}{{\bf a}}
\newcommand{\bb}{{\bf b}}
\newcommand{\pd}{\partial}

 \journalname{Journal of Mathematical Biology}
\begin{document}

\title{Entrainment in up and down states of neural populations: non-smooth and stochastic models \thanks{This work was supported by NSF DMS-1311755.}
}

\titlerunning{Population models of up and down states}        % if too long for running head

\author{Zachary McCleney         \and
        Zachary P. Kilpatrick
}

\authorrunning{McCleney \and Kilpatrick} % if too long for running head

\institute{Z. McCleney \and Z.P. Kilpatrick \at
              Department of Mathematics, University of Houston, Houston TX 77204 \\
              \email{zpkilpat@math.uh.edu}           %  \\
%             \emph{Present address:} of F. Author  %  if needed
           \and
           Z. McCleney \at
             \email{mccleneyzach@yahoo.com}
}

\date{Received: date / Accepted: date}
% The correct dates will be entered by the editor

\maketitle

\begin{abstract}

We study the impact of noise on a neural population rate model of up and down states. Up and down states are typically observed in neuronal networks as a slow oscillation, where the population switches between high and low firing rates \citep{sanchezvives00}. A neural population model with spike rate adaptation is used to model such slow oscillations, and the timescale of adaptation determines the oscillation period. Furthermore, the period depends non-monotonically on the background tonic input driving the population, having long periods for very weak and very strong stimuli. Using both linearization and fast-slow timescale separation methods, we can compute the phase sensitivity function of the slow oscillation. We find that the phase response is most strongly impacted by perturbations to the adaptation variable. Phase sensitivity functions can then be utilized to quantify the impact of noise on oscillating populations. Noise alters the period of oscillations by speeding up the rate of transition between the up and down states. When common noise is presented to two distinct populations, their transitions will eventually become entrained to one another through stochastic synchrony.

% \PACS{PACS code1 \and PACS code2 \and more}
% \subclass{MSC code1 \and MSC code2 \and more}
\end{abstract}

\section{Introduction}
\label{intro}

Cortical networks can generating a wide variety oscillatory rhythms with frequencies spanning five orders of magnitude \citep{buzsaki04}. Slow oscillatory activity (0.1-1Hz) has been observed {\em in vivo} during decreased periods of alertness, such as slow wave sleep and anesthesia \citep{steriade93}. Furthermore, such activity can be produced {\em in vitro} when bathing cortical slices in a medium with typical extracellular ion concentrations \citep{sanchezvives00}. A key feature of these slow oscillations is that they tend to be an alternating sequence of two bistable states, referred to as the {\em up} and {\em down} states. Up states in networks are characterized by high levels of firing activity, due to depolarization in single cells. Down states in networks typically appear quiescent, due to hyperpolarization in single cells. There is strong evidence that up states are neural circuit attractors, that emerge due to synaptic feedback \citep{cossart03}. This suggests up states may be spontaneous remnants of stimulus-induced persistent states utilized for working memory \citep{wang01} and other network computations \citep{major04}.

Several different cellular and synaptic mechanisms have been suggested to underlie the transitions between up and down states. One possibility is that the network is recurrently coupled with excitation, stabilizing both a quiescent and active state \citep{amit97,renart07}. Fluctuations due to probabilistic synapses, channel noise, and randomness in network connectivity can then lead to spontaneous transitions between the quiescent and active state \citep{bressloff10,litwinkumar12}. Alternatively, switches between low and high activity states may arise by some underlying systematic slow process. For instance, it has been shown that competition between recurrent excitation and the negative feedback produced by activity-dependent synaptic depression can lead to slow oscillations in firing rate whose timescale is set by the depression timescale \citep{bart05,kilpatrick10}. Excitatory-inhibitory networks with facilitation can produce slow oscillations, due to the slow facilitation of feedback inhibition that terminates the up state, the down state is then rekindled due to positive feedback from recurrent excitation \citep{melamed08}. These neural mechanisms utilize dynamic changes in the strength of neural architecture. However, \cite{compte03} proposed that single cell mechanisms can also shift network states between up and down states. The up state is maintained by strong recurrent excitation balanced by inhibition, and transitions to the down state occur due to a slow adaptation current. Once in the down state, the adaptation current is inactivated, and excitation reinitiates the up state. A similar mechanism has been utilized in models of perceptual rivalry, where dominance switches between two mutually inhibiting populations are due to the build up of a rate-based adaptation current \citep{laing02,morenobote07}.

In this paper, we utilize a rate-based model of an excitatory network with spike rate adaptation to explore the impact that noise perturbations have upon the relative phase and duration of slow oscillations. We find that, as in the spiking model studied by \cite{compte03}, the interplay between recurrent excitation and adaptation produces a slow oscillation in the firing rate of the network. In fact, for slow timescale adaptation currents, the oscillations evolve as fast switches between a low and high activity state, stable fixed points of the adaptation-free system. Since the timescale and slow dynamics of the oscillation are set by the adaptation current, we mainly focus on the impact of perturbation to the adaptation variable in our model. As we will show, perturbations of the activity variable have much lower impact on the oscillation phase. Introducing noise into the adaptation variable of the population model leads to a speeding up of the slow oscillation, due to early switching between the low and high activity state.

Another remarkable feature of slow oscillations, observed during slow-wave sleep and anesthesia, is that the up and down states tend to be synchronized across different regions of cortex and thalamus \citep{steriade93,massimini04}. Specifically, both the up and down states start near synchronously in cells located up to 12mm apart \citep{volgushev06}. Such remarkable coherence between distant network activity cannot be accomplished by single cell mechanisms, but require either long range network connectivity or some external signal forcing entrainment \citep{traub96,smeal10}. Activity tends to originate from several different foci in the network, quickly spreading across the rest of the network on a timescale orders of magnitude faster than the oscillation itself \citep{compte03,massimini04}. The fact that the onset of quiescence is fast and well synchronized means there must be either a rapid relay signal between all foci or there is some global signal cueing the down state. Rather than suggest a disynaptic relay, using long range excitation acting on local inhibition, we suggest that background noise can serve as a synchronizing global signal \citep{ermentrout08}. Noisy but correlated inputs have been shown to be capable of synchronizing uncoupled populations of  phase oscillators \citep{teramae04} as well as experimentally recorded cells in vitro \citep{galan06}. Here we will show correlated noise is a viable mechanisms for coordinating slow oscillations in distinct uncoupled neural populations.

The paper is organized as follows. We introduce the neural population model in section \ref{mod}, indicating the way external noise is incorporated into the model. In section \ref{periodicsoln}, we demonstrate the periodic solutions that emerge in the noise-free model, demonstrating it is possible to derive analytical expressions for the oscillation period in the case of steep firing rate functions. Then, in section \ref{prc_pws} we show how to derive phase sensitivity functions that describe how external perturbations to the periodic solution impact the asymptotic phase of the oscillation. As demonstrated, the impact of perturbations to the adaptation variable is much stronger than activity variable perturbations, especially for longer adaptation timescales. Thus, our studies of the impact of noise mainly focus on the effects of fluctuations in the adaptation variable. We find, in section \ref{1dnoise}, that adding noise to the adaptation variable leads to up and down state durations that are shorter and more balanced, so that the up and down state last for similar lengths of time. In section \ref{ssynch}, we demonstrate that slow oscillations in distinct populations can become entrained to one another when both populations are forced by the same common noise signal. This phenomenon is robust to the introduction of independent noise in each population, as we show in section \ref{indepnos}.

\section{Adaptive neural populations: deterministic and stochastic models}
\label{mod}

We begin by describing the models we will use to explore the impact of external perturbations on slow oscillations. Motivated by \cite{compte03}, we will focus on a neural population model with spike rate adaptation, akin to mutual inhibitory models used to study perceptual rivalry \citep{laing02,morenobote07}. 

{\bf Single population model.} In a single population, neural activity $u(t)$ receives negative feedback due to a subtractive spike rate adaptation term \citep{benda03}
\begin{subequations} \label{single}
\begin{align}
\dot{u}(t) &=  -u(t) + f(\alpha u(t)  - a(t) + I), \\
\tau \dot{a}(t) &= -a(t) + \phi u (t).
\end{align}
\end{subequations}
Here, $u$ represents the mean firing rate of the neural population with excitatory connection strength $\alpha$. The negative feedback variable $a$  is spike frequency adaptation with strength $\phi$ and time constant $\tau$. For some of our analysis we will utilize the assumption $\tau \gg 1$, based on the fact that many form for spike rate adaptation tend to be much slower than neural membrane time constants \citep{benda03}. The constant tonic drive $I$ initiates the high firing rate (up) state, and slow adaptation eventually attenuates activity to a low firing rate (down) state. Weak but positive drive $I>0$ is meant to model the presence of low spiking threshold cells that spontaneously fire, utilized as a mechanism for initiating the up state in \cite{compte03}. The firing rate function $f$ is monotone and saturating function such as the sigmoid
\begin{align}
f(x) = \frac{1}{1 + \e^{- \gamma x}}.  \label{sig}
\end{align}
Commonly, in studies of neural field models, the high gain limit of (\ref{sig}) is taken to yield the Heaviside firing rate function \citep{amari77,laing02}
\begin{eqnarray}
H(x) &=& \left\{ \begin{array}{cc} 1 : & x \geq 0, \\ 0 : & x < 0, \end{array} \right. \label{H}
\end{eqnarray}
which often allows for a more straightforward analytical study of model dynamics. We exploit this fact extensively in our study. Nonetheless, we have also carried out many numerical simulations of the model for a smooth firing rate function (\ref{sig}), and they correspond to the results we presentfor sufficiently high gain. Note, this form of adaptation is often referred to as {\em subtractive} negative feedback, as current is subtracted from the population input. Alternative models of slow neural population oscillations have employed short term synaptic depression \citep{tabak00,bart05,kilpatrick10}, a form of {\em divisive} negative feedback.

A primary concern of this work is the response of (\ref{single}) to external perturbations, acting on the activity $u$ and adaptation $a$ variables. To do so, we will use both an exact method and a linearization to identify the phase response curve of the limit cycle solutions to (\ref{single}). Understanding the susceptibility of limit cycles (\ref{single}) to inputs will help us understand ways in which noise will influence the frequency and regularity of oscillations.

{\bf Stochastic single population model.} Following our analysis of the noise-free system, we will consider how fluctuations influence oscillatory solutions to (\ref{single}). To do so, we will employ the following Langevin equation for (\ref{single}) forced by white noise
\begin{subequations}  \label{stochmod}
\begin{align}
\d u(t) &= \left[ - u(t) + f(\alpha u(t) - a(t) + I ) \right] \d t + \d \xi_u(t) \\
\d a(t) &= \left[ - a(t) + \phi u(t) \right] \d t/ \tau + \d \xi_a(t),
\end{align}
\end{subequations}
where we have introduced the independent Gaussian white noise processes $\xi_u(t)$ and $\xi_a(t)$ with zero mean $\langle \xi_u(t) \rangle = \langle \xi_a(t) \rangle = 0$ and variances $\langle \xi_u(t)^2 \rangle = \sigma_u^2 t$ and $\langle \xi_a(t)^2 \rangle = \sigma_a^2 t$. Extending our results concerning the phase response curve, we will explore how noise forcing impacts the statistics of the resulting stochastic oscillations in (\ref{stochmod}). In particular, since we find noise tends to impact the phase of the oscillation more strongly when applied to the adaptation variable, we will tend to focus on the case $\xi_u \equiv 0$.

{\bf Stochastic dual population model.} Finally, we will focus on how correlations in noise-forcing impact the coherence of two distinct uncoupled populations
\begin{subequations} \label{dual}
\begin{align}
\d u_1 &= \left[ - u_1(t) + f(\alpha u_1(t) - a_1(t) + I ) \right] \d t + \d \xi_{u} \\
\d a_1 &= \left[ - a_1(t) + u_1(t) \right] \d t / \tau + \d \xi_{a} \\
\d u_2 &= \left[ -u_2 (t) + f(\alpha u_2(t) - a_2(t) + I ) \right] \d t + \d \xi_{u} \\
\d a_2 &= \left[ -a_2 (t) + u_2 (t) \right] \d t / \tau + \d \xi_{a}.
\end{align}
\end{subequations}
Thus, the system (\ref{dual}) describes the dynamics of two distinct neural populations $u_1$ and $u_2$, with inputs $I$. Our main interest lies in the impact the noise terms have upon the phase relationship between the two systems' states. In this version of the model, noise to the activity variables $\xi_u$ is totally correlated, as is noise to the adaptation variables $\xi_a$. Thus, all means are zero and $\langle \xi_{u}^2 (t) \rangle = \sigma_u^2 = \Db_{11} t$. Furthermore, $\langle \xi_{a}^2 (t) \rangle = \sigma_a^2 t = \Db_{22} t$. For this study, we assume there are no correlations between activity and adaptation noise, so $\langle \xi_u(t) \xi_{a} (t) \rangle = 0$. A more general version of the model (\ref{dual}) would consider the possibility of independent noise in each population
\begin{subequations} \label{dualind}
\begin{align}
\d u_1 &= \left[ - u_1(t) + f(\alpha u_1(t) - a_1(t) + I ) \right] \d t + \chi_u \d \xi_{uc} + \sqrt{1 - \chi_u^2} \d \xi_{u1} \\
\d a_1 &= \left[ - a_1(t) + u_1(t) \right] \d t / \tau +  \chi_a \d \xi_{ac} + \sqrt{1 - \chi_a^2} \d \xi_{a1}  \\
\d u_2 &= \left[ -u_2 (t) + f(\alpha u_2(t) - a_2(t) + I ) \right] \d t +  \chi_u \d \xi_{uc} + \sqrt{1 - \chi_u^2} \d \xi_{u2}  \\
\d a_2 &= \left[ -a_2 (t) + u_2 (t) \right] \d t / \tau +  \chi_a \d \xi_{ac} + \sqrt{1 - \chi_a^2} \d \xi_{a2} .
\end{align}
\end{subequations}
Noise terms all have zero mean and variances defined $\langle \xi_{uj}^2 (t) \rangle = \sigma_{uj}^2 t = \Db_{uj} t$ and $\langle \xi_{aj}^2 (t) \rangle = \sigma_{aj}^2 t= \Db_{aj} t$ ($j=1,2,c$). To ease calculations, we take $\Db_{u1} = \Db_{u2} \equiv \Db_{ul} = \sigma_u^2$ and $\Db_{a1} = \Db_{a2} \equiv \Db_{al} = \sigma_a^2$. The degree of noise correlation between populations is controlled by the parameters $\chi_u$ and $\chi_a$, so in the limit $\chi_{u,a} \to 1$, the model (\ref{dualind}) becomes (\ref{dual}).

\section{Periodic solutions of a single population}
\label{periodicsoln}

\begin{figure}
\begin{center} \includegraphics[width=6cm]{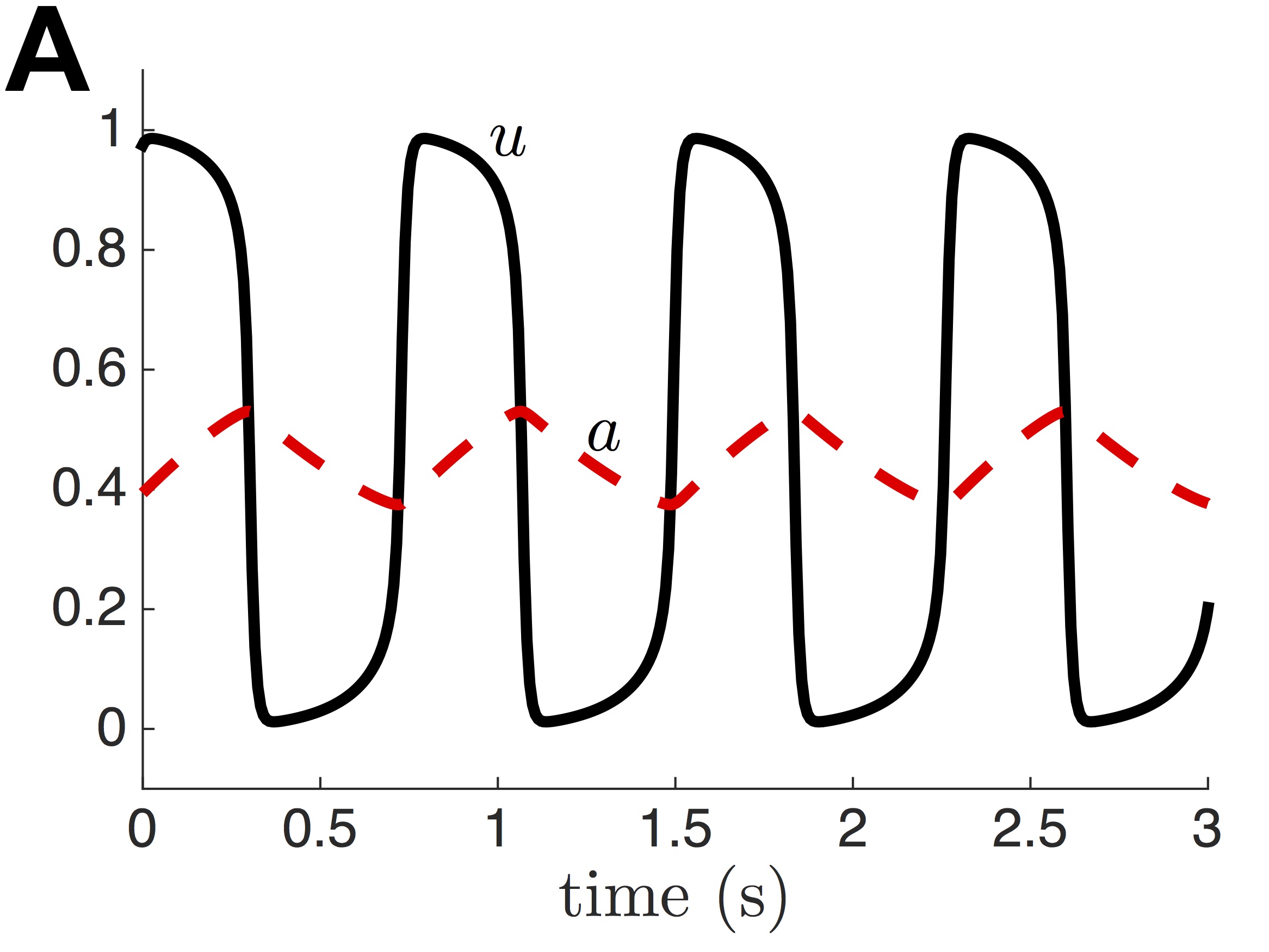} \includegraphics[width=6cm]{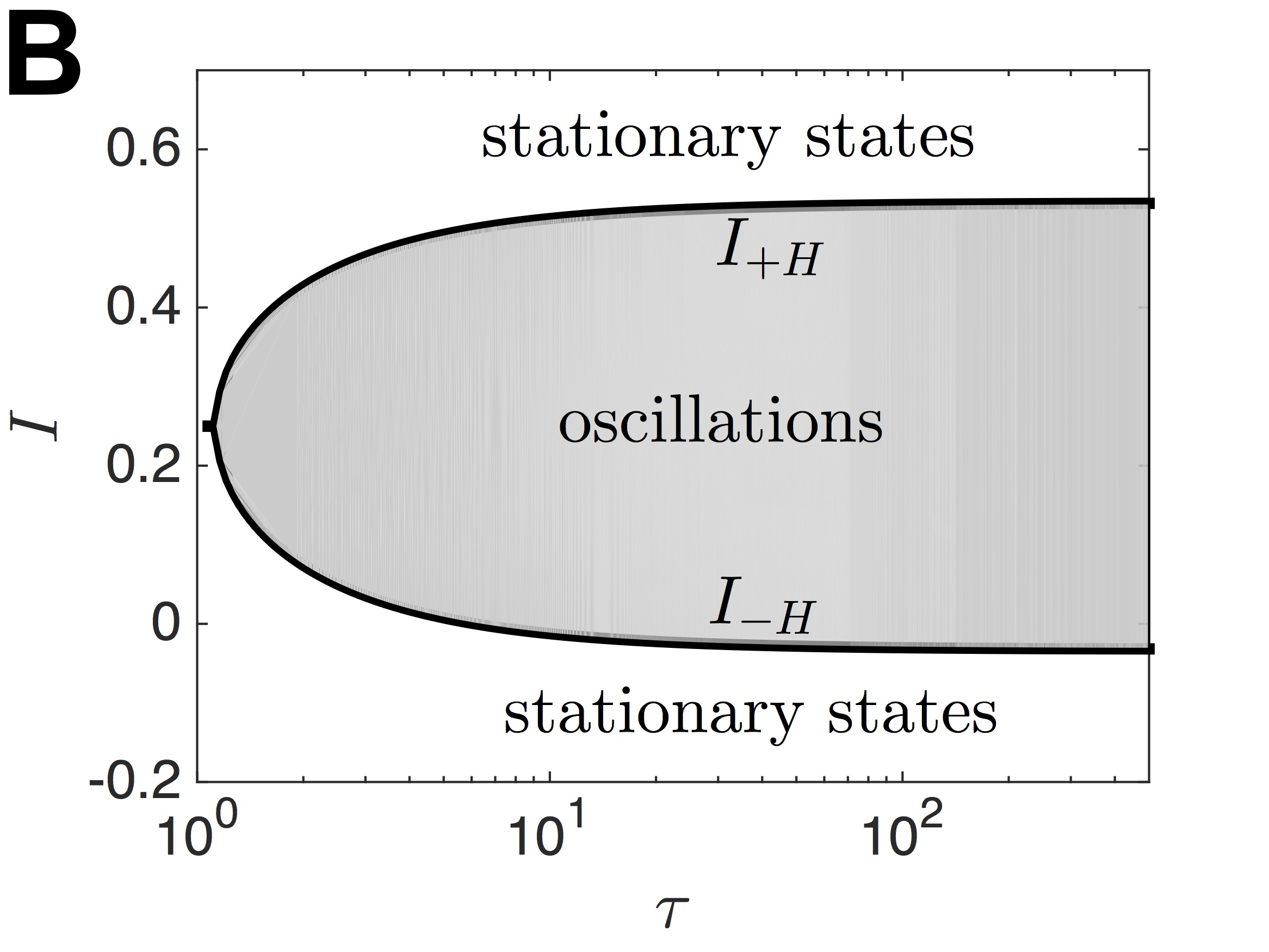} \\  \includegraphics[width=6cm]{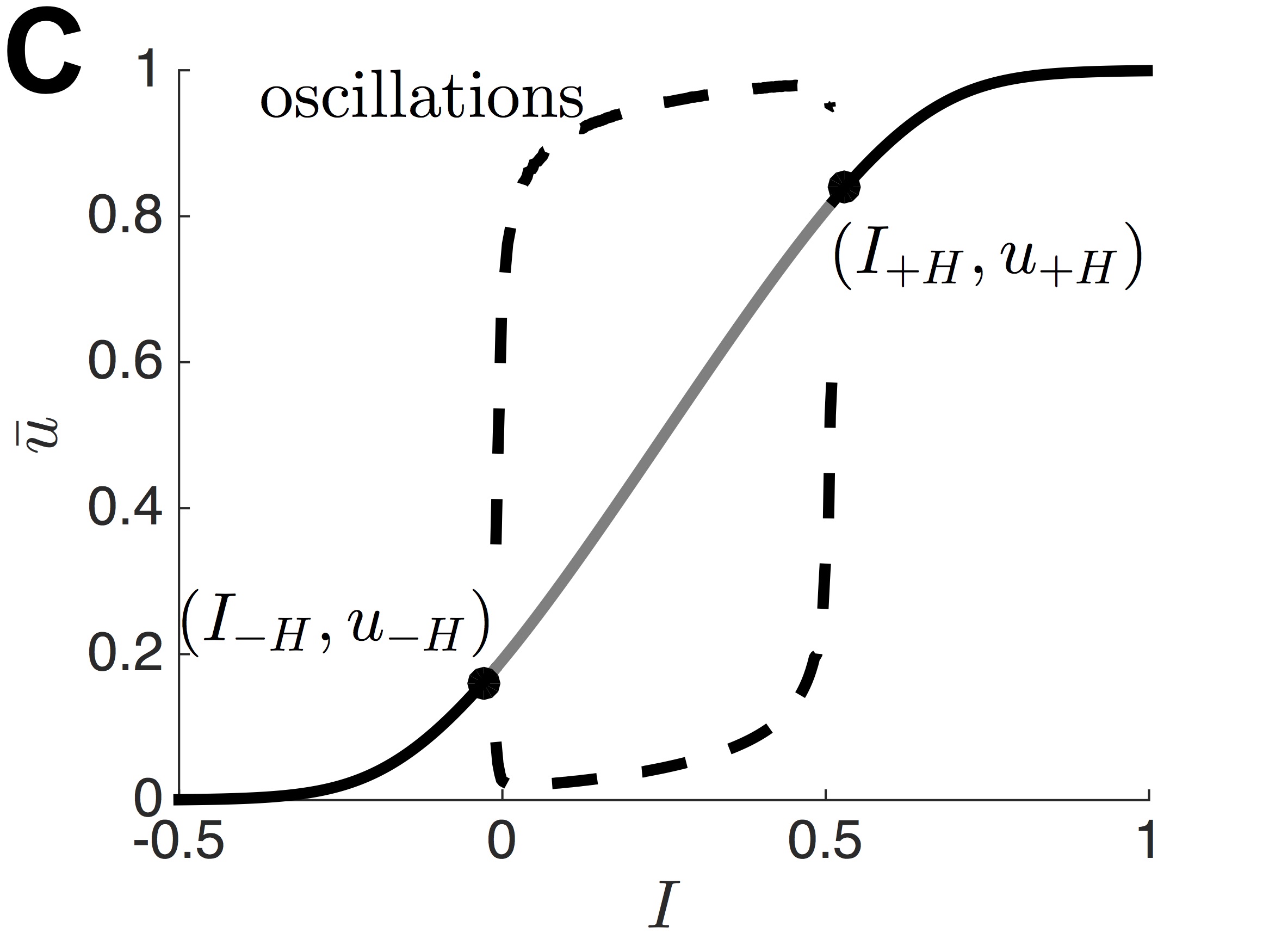} \includegraphics[width=6cm]{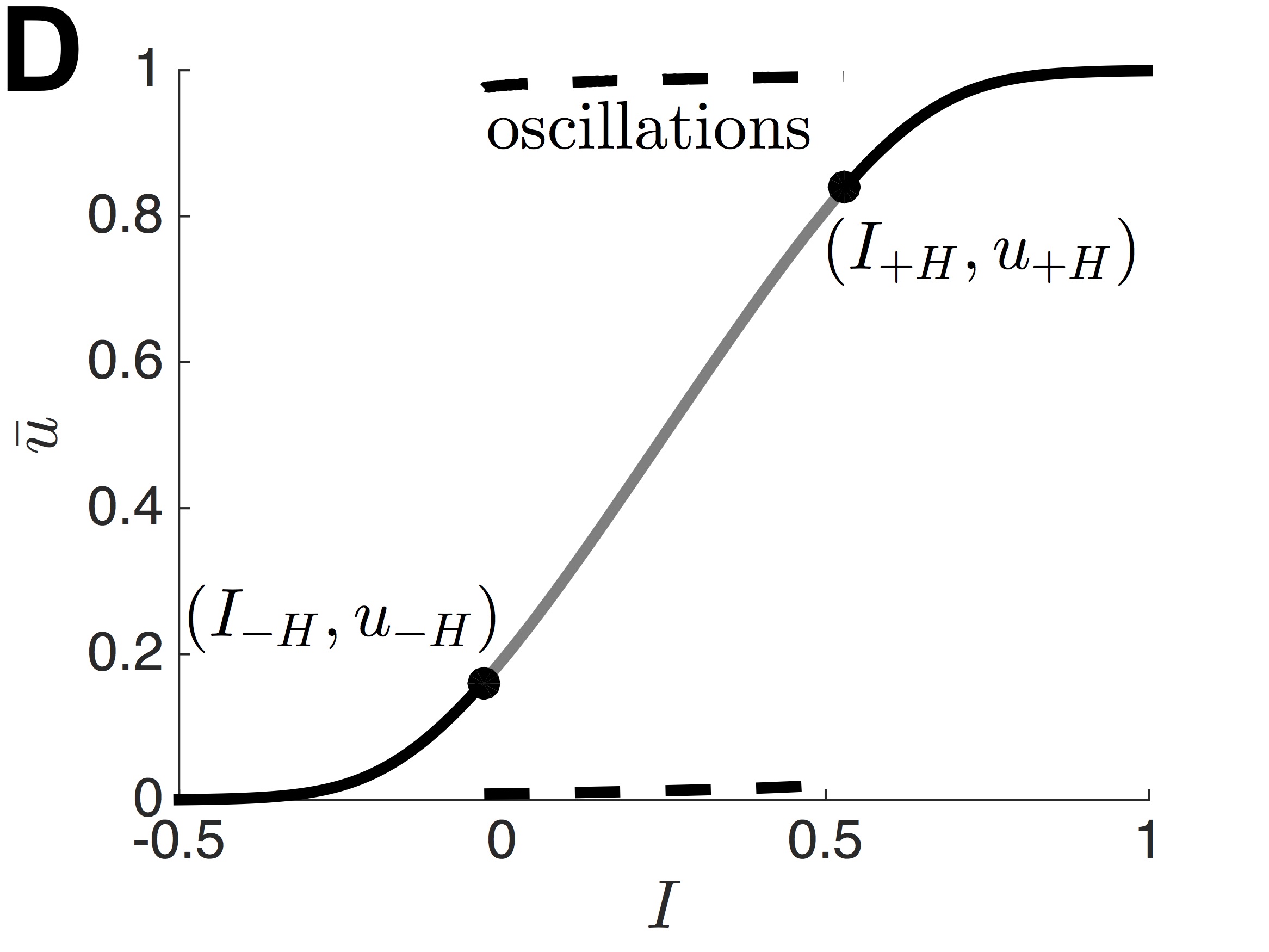} \end{center}
\caption{Single adapting neural population (\ref{single}) generates slow oscillations. ({\bf A}) Numerical simulation of (\ref{single}) for adaptation timescale $\tau = 100$ (1s) and input $I=0.2$. ({\bf B}) Partitioning of $(\tau, I)$ parameter space shows the range of inputs $I$ leading to oscillations expands as the adaptation timescale $\tau$ is increased, according to (\ref{Ihopf}). ({\bf C},{\bf D}) Bifurcation diagram showing supercritical ($I_{-H}$) and subcritical ($I_{+H}$) Hopf bifurcations that arise as the input is increased for ({\bf C}) $\tau=10$ and ({\bf D}) $\tau=100$.  Firing rate function is sigmoidal (\ref{sig}). Other parameters are $\phi = 1$, $\alpha = 0.5$, and $\gamma = 15$}
\label{singlefp}
\end{figure}

We begin by studying periodic solutions of the single population system (\ref{single}), as demonstrated in Fig. \ref{singlefp}{\bf A}. First, we note that for firing rate functions $f$ with finite gain, we can identify the emergence of oscillations by analyzing the stability of the equilibria of (\ref{single}). That is, we assume $(\dot{u}, \dot{a}) = (0,0)$, so the system becomes
\begin{align*}
\bar{u} &= f( \alpha \bar{u} - \bar{a} + I), \\
\bar{a} &= \phi \bar{u},
\end{align*}
which can be reduced to the single equation
\begin{align}
\bar{u} &= f((\alpha - \phi ) \bar{u} + I) = g(\bar{u}).  \label{sfpeqn}
\end{align}
Roots of (\ref{sfpeqn}), defining fixed points of (\ref{single}) are plotted as a function of the input $I$ in Fig. \ref{singlefp}{\bf C},{\bf D}. Utilizing the sigmoidal firing rate function $f$ given by (\ref{sig}), we can show that there will be a single fixed point as long as $\phi > \alpha$. In this case, we can compute
\begin{align*}
\frac{\d g(\bar{u})}{\d \bar{u}} = - (\phi - \alpha) f'((\alpha - \phi ) \bar{u} + I) = - \frac{(\phi - \alpha) \e^{- \gamma ((\alpha - \phi ) \bar{u} + I)}}{\left( 1 + \e^{- \gamma ((\alpha - \phi ) \bar{u} + I)} \right)^2} <0.
\end{align*}
Since $\bar{u}$ is monotone increasing, then $\bar{u} - g(\bar{u})$ is monotone increasing. Further, noting $\lim_{\bar{u} \to \pm \infty} \left[ \bar{u} - g(\bar{u}) \right] = \pm \infty$, it is clear $\bar{u} - g(\bar{u})$ crosses zero once, so (\ref{sfpeqn}) has a single root when $\phi > \alpha$. Stability of this equilibrium is given by the eigenvalues of the associated Jacobian
\begin{align*}
J(\bar{u}, \bar{a}) = \left( \begin{array}{cc} -1 + \alpha f'((\alpha - \phi) \bar{u} + I ) & - f'((\alpha - \phi) \bar{u} +I) \\  \phi / \tau & -1/ \tau   \end{array} \right).
\end{align*}
We note that the sigmoid (\ref{sig}) satisfies the Ricatti equation $f' = \gamma f (1 - f)$, so we can use (\ref{sfpeqn}) to write
\begin{align*}
J(\bar{u}, \bar{a}) = \left( \begin{array}{cc} -1 + \alpha \gamma \bar{u} (1- \bar{u}) & - \gamma \bar{u} (1- \bar{u})  \\ \phi / \tau & -1 / \tau   \end{array} \right).
\end{align*}
Oscillations arise when stable spiral equilibria destabilize through a Hopf bifurcation. Hopf bifurcations will occur when complex eigenvalues associated with fixed points $(\bar{u}, \bar{a})$ cross from the left to the right half plane. We can require this with the pair of expression: ${\rm tr}(J) = 0$ and ${\rm tr}(J)^2 < 4 {\rm det} (J)$. Thus, a necessary condition for the Hopf bifurcation point is that the equilibrium value $\bar{u}$ satisfy
\begin{align*}
\alpha \gamma \bar{u} ( 1- \bar{u}) = 1 + 1 / \tau.
\end{align*}
Solving this for $\bar{u}$ yields
\begin{align}
\bar{u}_{\pm H} = \frac{1}{2} \left[ 1 \pm \sqrt{1 - 4 \chi} \right],  \hspace{4mm} \chi = \frac{1 + 1 / \tau}{\alpha \gamma}.  \label{uhopf}
\end{align}
Thus, Hopf bifurcations will only occur when the timescale of adaptation is sufficiently large $\tau > \left[ \alpha \gamma / 4 - 1\right]^{-1}$. Plugging the formula (\ref{uhopf}) back into the fixed point equation (\ref{sfpeqn}) and solving for the input $I$, we can parameterize Hopf bifurcation curves based upon the equation
\begin{align}
I_{\pm H} = \frac{1}{\gamma} \ln \left[ \frac{\bar{u}_{\pm H}}{1 - \bar{u}_{\pm H}} \right] - (\alpha - \phi ) \bar{u}_{\pm H},  \label{Ihopf}
\end{align}
along with the additional condition ${\rm tr}(J)^2 < 4 {\rm det} (J)$ which becomes
\begin{align}
\frac{4}{\tau^2} < \frac{4 \phi}{\alpha \tau^2} + \frac{4 \phi}{\alpha \tau},  \label{hopfineq}
\end{align}
which will always hold as long as $\phi > \alpha$. We partition the parameter space $(\tau, I)$ using our formula for the Hopf curve (\ref{Ihopf}) in Fig. \ref{singlefp}{\bf B}. As demonstrated, there tend to be either two or zero Hopf bifurcation points for a given timescale $\tau$, and the coalescence of the two Hopf points is given by the point where $\tau  = \left[ \alpha \gamma / 4 - 1\right]^{-1}$. 

In the limit of slow adaptation $\tau \gg 1$, we can separate the timescales of the activity $u$ and adaptation $a$ variables, finding $u$ will equilibrate according to the equation
\begin{align}
\hat{u}(t) = f(\alpha \hat{u}(t) - a(t) + I), \label{uqss}
\end{align}
and subsequently $a$ will slowly evolve according to the equation
\begin{align}
\dot{a}(t) = \left[ \phi \hat{u}(t) - a \right] / \tau. \label{aqss}
\end{align}
We always have an implicit formula for $\hat{u}(t)$ in terms of $a(t)$, so the dynamics will tend to slowly evolve along the direction of the $a$ variable. This demonstrates why periodic solutions to (\ref{single}) are comprised of a slow rise and decay phase of $a$, punctuated by fast excursions in the activity variable $u$. In general, it is not straightforward to analytically treat the pair of equations (\ref{uqss}) and (\ref{aqss}), but we will show how computing solutions of the singular system becomes straightforward when we take the high gain limit $\gamma \to \infty$.

\begin{figure}
\begin{center} \includegraphics[width=6cm]{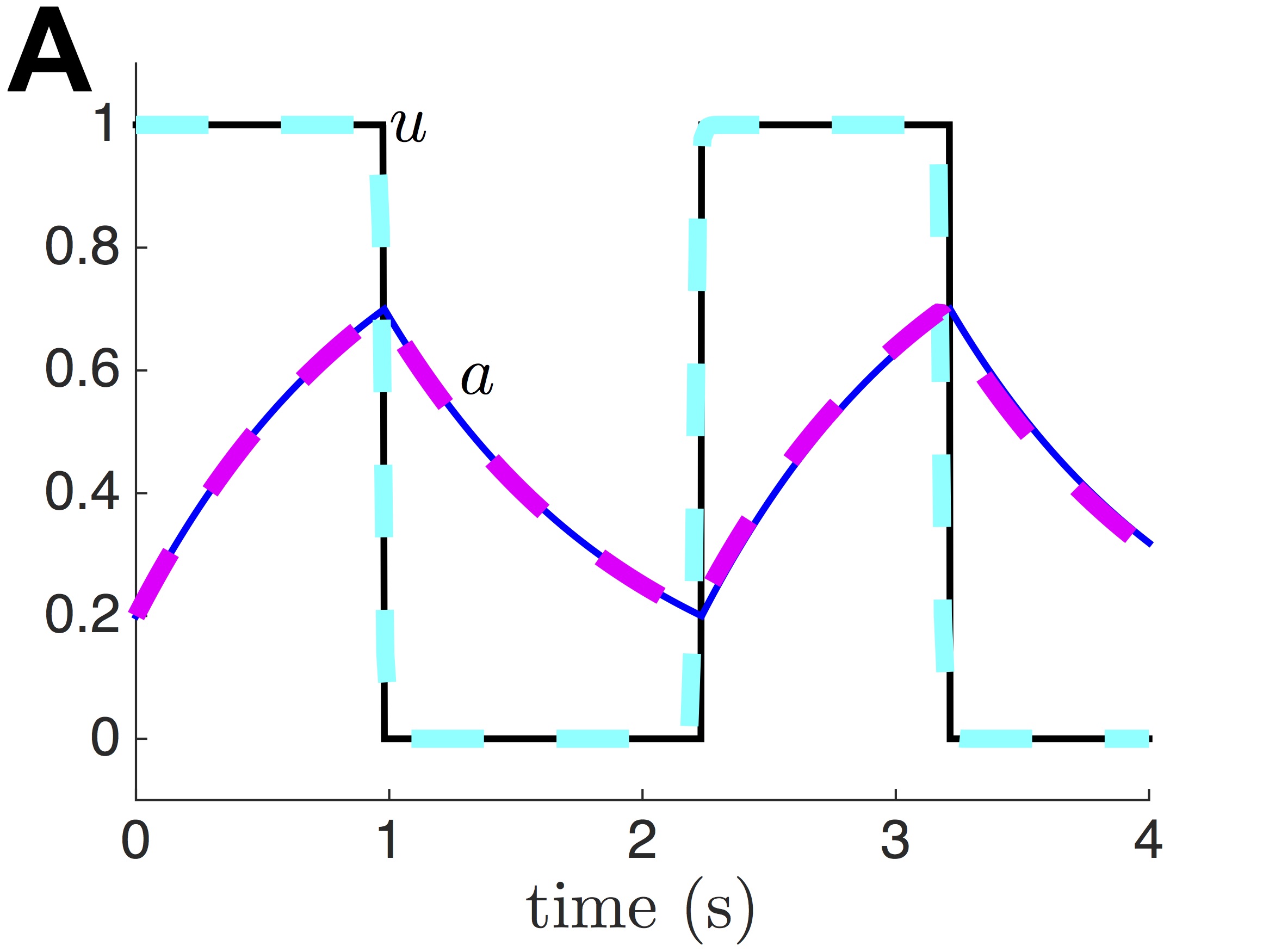} \includegraphics[width=6cm]{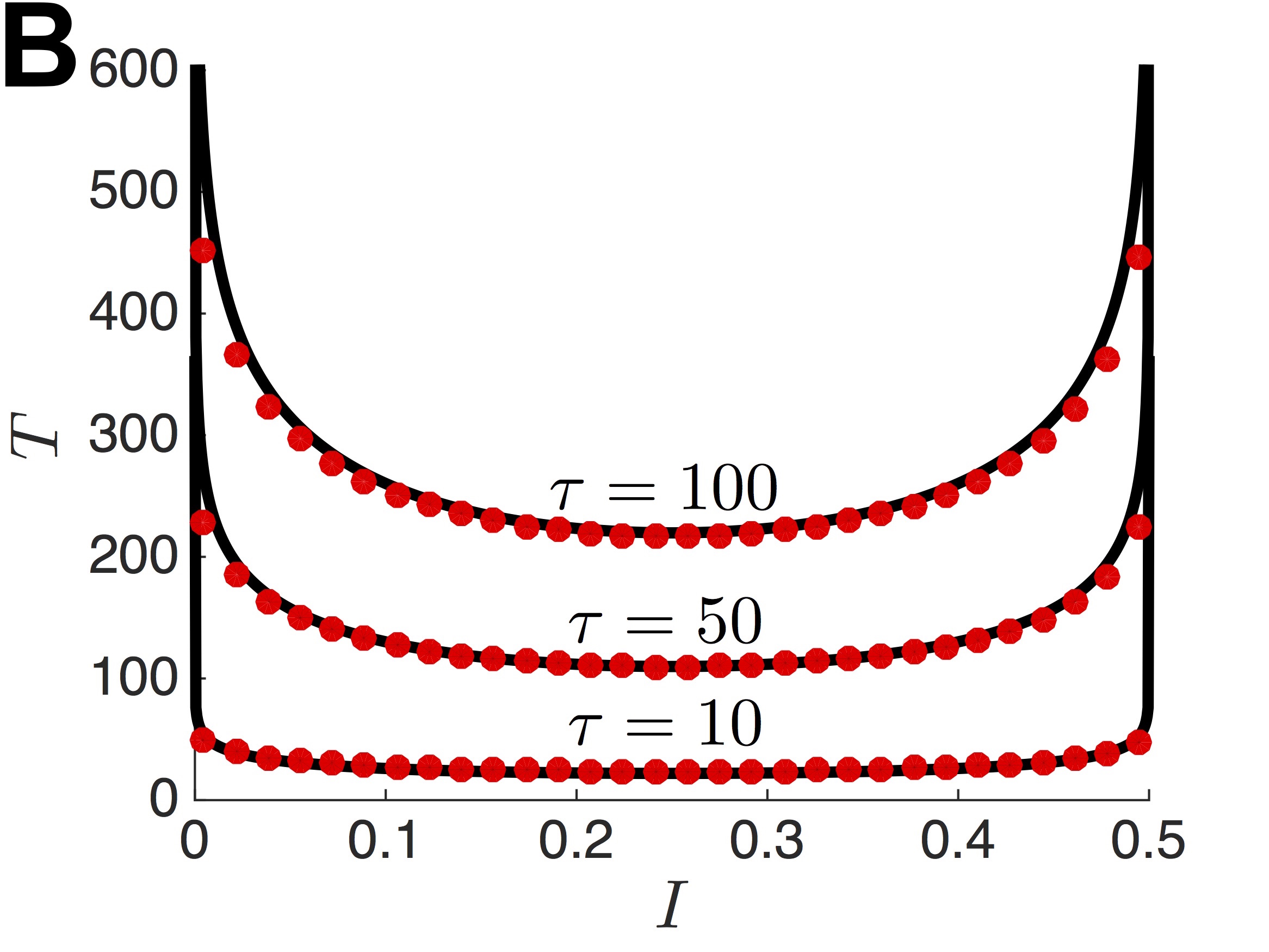} \end{center}
\caption{Analytical approximations to periodic solutions of (\ref{single}) with a Heaviside firing rate function (\ref{H}). ({\bf A}) Numerical simulation (solid lines) of the periodic solution is well approximated by the analytical approximation (dashed lines) given by (\ref{Hpersol}) when $I=0.2$ and $\tau = 100$. ({\bf B}) The period of the oscillation $T$ computed from numerical simulations (dots) is accurately approximated by the analytical formula (solid lines) given by (\ref{singper}). Other parameters are $\alpha = 0.5$ and $\phi = 1$.}
\label{Hperplots}
\end{figure}

Having established the existence of periodic solutions to (\ref{single}) in the case of sigmoid firing rates (\ref{sig}), we now explore the system in the high gain limit $\gamma \to \infty$ whereby the firing rate function becomes a Heaviside (\ref{H}). In this case, fixed points $(\bar{u},\bar{a})$ satisfy the equations
\begin{align*}
\bar{u} = H( ( \alpha - \phi ) \bar{u} + I ) = \left\{ \begin{array}{ll} 1 \ &: \bar{u} < I/(\phi - \alpha), \\ 0 \  &: \bar{u} > I/(\phi - \alpha), \end{array} \right.
\end{align*}
and $\bar{a} = \phi \bar{u}$. Thus, assuming $\phi > \alpha$, then $\bar{u}=0$ when $I<0$ and $\bar{u} = 1$ when $I> (\phi - \alpha)$. In both cases, the fixed points are linearly stable. When $0 < I < (\phi - \alpha)$, there are no fixed points and we expect to find oscillatory solutions. Assuming $\tau \gg 1$, we can exploit a separation of timescales to identify the shape and period of these limit cycles. To begin, we note that on fast timescales
\begin{align*}
\dot{u}(t) &= - u(t) + H(I + \alpha u(t) - a_0),
\end{align*}
where $a_0$ is a quasi steady state. On slow timescales on the order of $\tau$, then $u(t)$ quickly equilibrates and
\begin{subequations} \label{fastsub}
\begin{align}
u(t) &= H(I + \alpha u(t) - a(t)), \\
\tau \dot{a}(t) &= -a(t) + \phi u(t).
\end{align}
\end{subequations}
Periodic solutions to (\ref{single}) must obey the condition $(u(t),a(t)) = (u(t+nT),a(t+nT))$ for $t \in [0,T]$ and $n \in {\mathbb Z}$, so we focus on the domain $t \in [0,T]$. Examining (\ref{fastsub}), we can see oscillations in (\ref{single}) involve switches between $u(t) \approx 1$ and $u(t) \approx 0$. We translate time so that $u(t) \approx 1$ on $t \in [0,T_1)$ and $u(t) \approx 0$ on $t \in [T_1,T)$. Subsequently, this means for $t \in [0,T_1]$ the system (\ref{fastsub}) becomes $u \equiv 1$ and $\tau \dot{a} = -a + \phi$ so $a(t) = \phi - (\phi - I) \e^{-t/\tau}$. We know $a(0)=I$ because $u(0^-) \equiv 0$ in (\ref{fastsub}), and the argument of $H(x)$ must have crossed zero at $t=0$. In a similar way, we find on $t \in [T_1,T)$ that $u \equiv 0$ and $a(t) = (I+ \alpha) \e^{-(t-T_1)/\tau}$. Using the conditions $a(T_1) = I+ \alpha$ and $a(T) = I$, we find that the rise time of the adaptation variable (or the duration of the {\em up} state) is 
\begin{align*}
T_1 &= \tau \ln \left[ \frac{\phi - I}{\phi - \alpha - I} \right],
\end{align*}
and the decay time (or the duration of the {\em down} state) is
\begin{align*}
T_2 &= \tau \ln \left[ \frac{I+\alpha}{I} \right],
\end{align*}
and the total period of the oscillation is
\begin{align}
T &= \tau \ln \left[ \frac{(I+ \alpha) (\phi - I)}{I(\phi - \alpha - I)}  \right]. \label{singper}
\end{align} 
Thus, approximate periodic solutions to (\ref{single}) in the case of a Heaviside firing rate (\ref{H}) take the form
\begin{subequations} \label{Hpersol}
\begin{align}
u(t) &= \left\{ \begin{array}{ll} 1 \ & : t \in [0,T_1), \\ 0 \ & : t \in [T_1,T], \end{array} \right. \\
a(t) &= \left\{ \begin{array}{ll} \phi - (\phi - I) \e^{-t/\tau}  \ & : t \in [0,T_1), \\ (I+ \alpha) \e^{-(t-T_1)/\tau} \ & : t \in [T_1,T]. \end{array} \right.
\end{align}
\end{subequations}
We demonstrate the accuracy of the approximation (\ref{Hpersol}) in Fig. \ref{Hperplots}{\bf A}. Furthermore, we show that relationship between the period $T$ and model parameters is well captured by the formula (\ref{singper}). Notice there is a non-monotonic relationship between the period $T$ and the input $I$. We can understand this further by noting that the rise time $T_1$ of the adaptation variable $a$ increases monotonically with input
\begin{align*}
\frac{\d T_1}{\d I} = \frac{\tau \alpha}{(\phi - I)(\phi - \alpha - I )} > 0,
\end{align*}
when $0 < I < (\phi - \alpha)$. Furthermore, the decay time $T_2$ of the adaptation variable $a$ decreases monotonically with input
\begin{align*}
\frac{\d T_2}{\d I} = - \frac{\tau \alpha}{I(I+ \alpha)} < 0,
\end{align*}
when $0 < I < (\phi - \alpha)$. Thus, as $I \to 0^+$, the slow oscillation's period $T$ is dominated by very long decay times $T_2 \gg 1$ and as $I \to (\phi - \alpha)^-$, it is dominated by very long rise times $T_1 \gg 1$. We can identify the minimal period as a function of the input $I$ by finding the critical point of $T(I)$. To do so, we differentiate and simplify
\begin{align*}
\frac{\d T}{\d I} = - \frac{\tau \alpha \phi (2 I - (\phi - \alpha))}{I(I+\alpha)(\phi - I)(\phi - \alpha - I)},
\end{align*}
so the critical point of $T(I)$ is $I_{crit} = (\phi - \alpha)/2$, which corresponds to the minimal value of the period $T_{min}(I) = 2 \tau \ln \left[ (\phi + \alpha)/(\phi - \alpha) \right]$ as pictured in Fig. \ref{Hperplots}{\bf B}.

\section{Phase response curves}
\label{prc_pws}

We can further understand the dynamics of the slow oscillations in (\ref{single}) by computing phase response curves for both the case of a sigmoidal firing rate (\ref{sig}) and the Heaviside firing rate (\ref{H}). As we will show, perturbations of the activity variable $u$ have decreasing impact as the timescale of adaptation $\tau$ and the gain $\gamma$ of the firing rate are increased. Perturbations of the adaptation variable $a$ tend to dominate the resulting dynamics, as it is the evolution of this slow variable that primarily determines the phase of the oscillation.

\begin{figure}
\begin{center} \includegraphics[width=4cm]{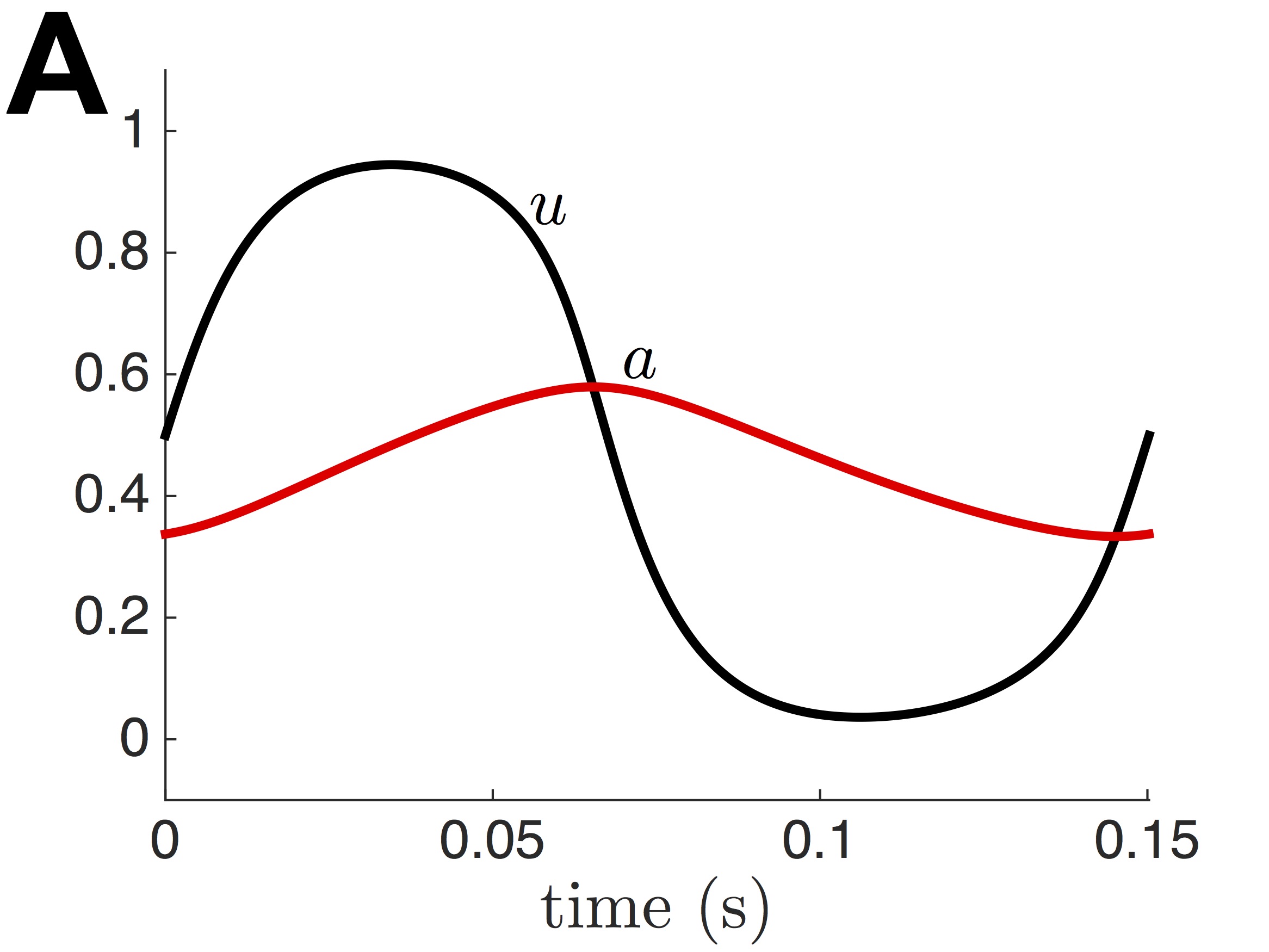} \includegraphics[width=4cm]{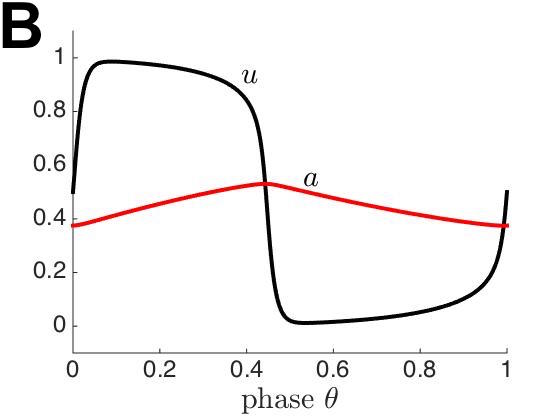} \includegraphics[width=4cm]{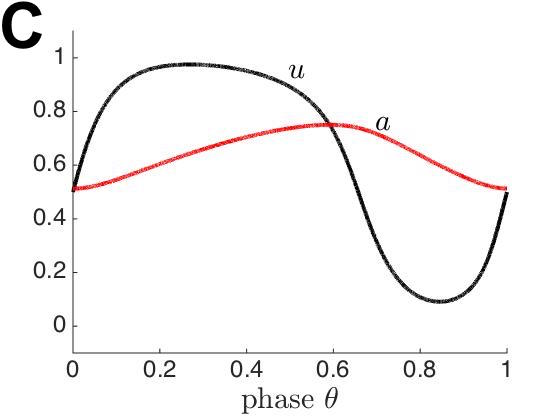} \\ \includegraphics[width=4cm]{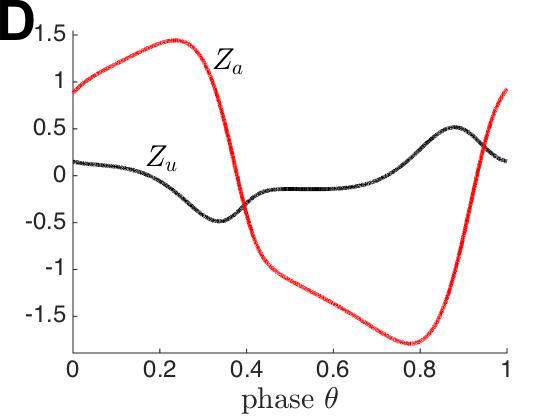} \includegraphics[width=4cm]{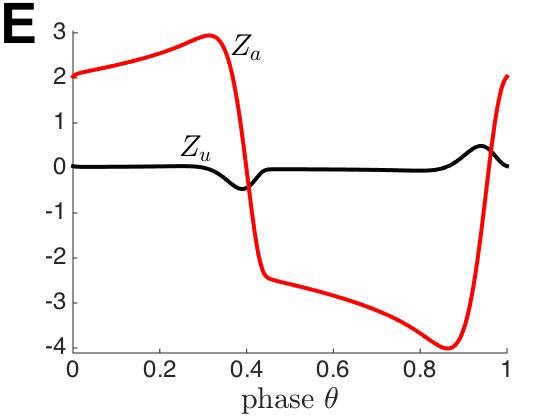} \includegraphics[width=4cm]{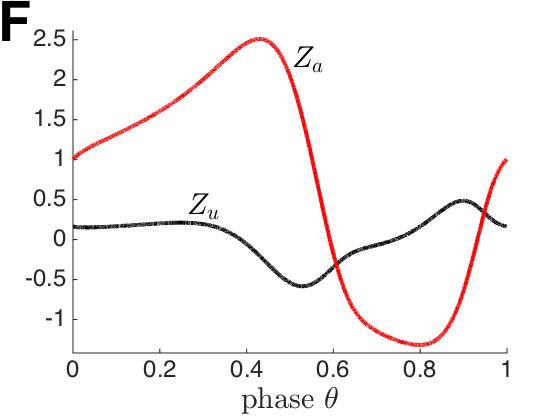} \end{center}
\caption{({\bf A}, {\bf B}, {\bf C}) Periodic solution $(u,a)$ and  ({\bf D}, {\bf E}, {\bf F}) phase sensitivity function $(Z_u,Z_a)$ of (\ref{single}) plotted as a function of phase $\theta = t/T$ for a sigmoidal firing rate function (\ref{sig}). ({\bf A},{\bf D}) For shorter adaptation timescale $\tau = 10$ and input $I=0.2$, the activity variable $u$ has a more rounded trajectory, so perturbations to activity influence the oscillation phase more (note size of lobes in on $Z_u$ in ({\bf D}). ({\bf B},{\bf E}) As the adaptation timescale is increased to $\tau =100$, with $I=0.2$, the influence of perturbations to the activity variable decrease (compare lobes of $Z_u$ to those in ({\bf D}). Perturbations of the adaptation variable influence the phase more strongly as shown by the change in the relative amplitude of $Z_a$. ({\bf C},{\bf F}) Increasing the input $I=0.4$, with $\tau = 10$, increases the relative duration of the rise time of $a$. As a result, there is a wider region where perturbations to $a$ advance the phase. Other parameters are $\alpha = 0.5$, $\phi =1$, and $\gamma = 15$.}
\label{prcsig}
\end{figure}

To begin, we derive a general formula that linearly approximates the influence of small perturbations on limit cycle solutions $(u_0(t), a_0(t))$ to (\ref{single}). Essentially, we utilize the fact that solutions $\Z (t)$ to the adjoint equation associated with linearization about the limit cycle solution $(u_0(t), a_0(t))$ provide a complete description of how infinitesimal perturbations of the limit cycle impact its phase \citep{ermentrout96,brown04}. To start, we note that
\begin{align*}
{\mc L} \left( \begin{array}{c} u_1 \\ a_1 \end{array} \right) = \left( \begin{array}{l} \dot{u_1}  +u_1 - \alpha f'(\alpha u_0 - a_0 + I) u_1 + f'(\alpha u_0 - a_0 + I) a_1, \\
\dot{a_1} - \phi u_1/\tau + a_1/\tau \end{array} \right) = \left( \begin{array}{c} 0 \\ 0 \end{array} \right),
\end{align*}
is the linearization of (\ref{single}) about the limit cycle $(u_0(t), a_0(t))$. Defining the inner product on $T$-periodic functions in $\R^2$ as $\langle F(t), G(t) \rangle = \int_0^T F(t) \cdot G(t) \d t$, we can find the adjoint operator ${\mc L}^*$ by noting it satisfies $\langle F, {\mc L}G \rangle = \langle {\mc L}^*F, G \rangle$ for all $L^2$ integrable vector functions $F,G$. We can then compute
\begin{align}
{\mc L}^* \left( \begin{array}{c} v \\ b \end{array} \right) = \left( \begin{array}{l} - \dot{v} + v - \alpha f'(\alpha u_0 - a_0 + I) v - \phi b / \tau \\ -\dot{b} + f'(\alpha u_0 - a_0 + I) v +b/ \tau \end{array} \right).
\end{align}
It can be shown that the null space of ${\mc L}^*$ describes the response of the phase of the limit cycle $(u_0(t),a_0(t))$ to infinitesimal perturbations \citep{brown04}. Note that if $(u_0(t),a_0(t))$ is a stable limit cycle then the nullspace of ${\mc L}$ is spanned by scalar multiples of $(u_0'(t),a_0'(t))$. Furthermore, appropriate normalization requires that $\Z(t) \cdot (u_0'(t),a_0'(t)) = 1$ along with ${\mc L}^* \Z = 0$ \citep{ermentrout96}. To numerically compute $\Z(t) = (Z_u(t), Z_a(t))$, we thus integrate the system
\begin{subequations}  \label{adjeqns}
\begin{align}
\dot{Z}_u(t) &= Z_u(t) - \alpha f'(\alpha u_0(t) - a_0(t) + I) Z_u(t) - \phi Z_a(t) / \tau, \\
\dot{Z}_a(t) &= f'(\alpha u_0(t) - a_0(t) + I) Z_u(t) - Z_a(t)/ \tau,
\end{align}
\end{subequations}
backward in time, taking the long time limit to find $(Z_u(t), Z_a(t))$ on $t \in [0,T]$, and normalizing $\langle (Z_u(t), Z_a(t)),(u_0'(t),a_0'(t)) \rangle = 1$ by rescaling appropriately. We demonstrate this result in Fig. \ref{prcsig}, showing the relationship between the shape and relative amplitude of the phase sensitivity functions $(Z_u,Z_a)$ and the parameters. Notably, perturbing the activity variable $u$ become less influential as the timescale of adaptation $\tau$ is increased ($Z_u$). Furthermore, there is a sharper transition between phase advance and phase delay region of the adaptation phase response ($Z_a$) for larger timescales $\tau$.

In addition to a general formula for the phase sensitivity functions $(Z_u(t),Z_a(t))$, we can derive an amplitude dependent formula for the response of limit cycle solutions $(u_0(t),a_0(t))$ of (\ref{single}) with a Heaviside firing rate (\ref{H}), assuming $\tau \gg 1$. In this case, we utilize the formula for the period (\ref{singper}) and limit cycle (\ref{Hpersol}), derived using a separation of timescales assumption. Then, we can compute the change to the variables $(u,a)$ as a result of a perturbation $(\delta_u, \delta_a)$, which we denote $(u_0(t),a_0(t)) \stackrel{(\delta_u, \delta_a)}{\longmapsto} (\tilde{u}_0(t), \tilde{a}_0(t))$. We are primarily interested in how the relative time in the limit cycle is altered by a perturbation $\delta_u$ - how much closer or further the limit cycle is to the end of the period $T$ after being perturbed. We can readily determine this by first inverting the formula we have for $(u_0(t),a_0(t))$, given by (\ref{Hpersol}), to see how this value determines the time $t_0$ along the limit cycle
\begin{align}
t_0 (u_0,a_0)  = \left\{ \begin{array}{cl} \tau \ln \left[ (\phi - I)/(\phi - a_0) \right] & : u_0 = 1, \\ \tau \ln \left[ (\phi - I)(I+\alpha)/a_0/(\phi - \alpha - I) \right] & : u_0 = 0. \end{array} \right.  \label{invertime}
\end{align}
Using this formula, we can now map the value $(\tilde{u}_0,\tilde{a}_0)$ to an associated updated relative time $t_0$ along the oscillation.

\begin{figure}
\begin{center} \includegraphics[width=4cm]{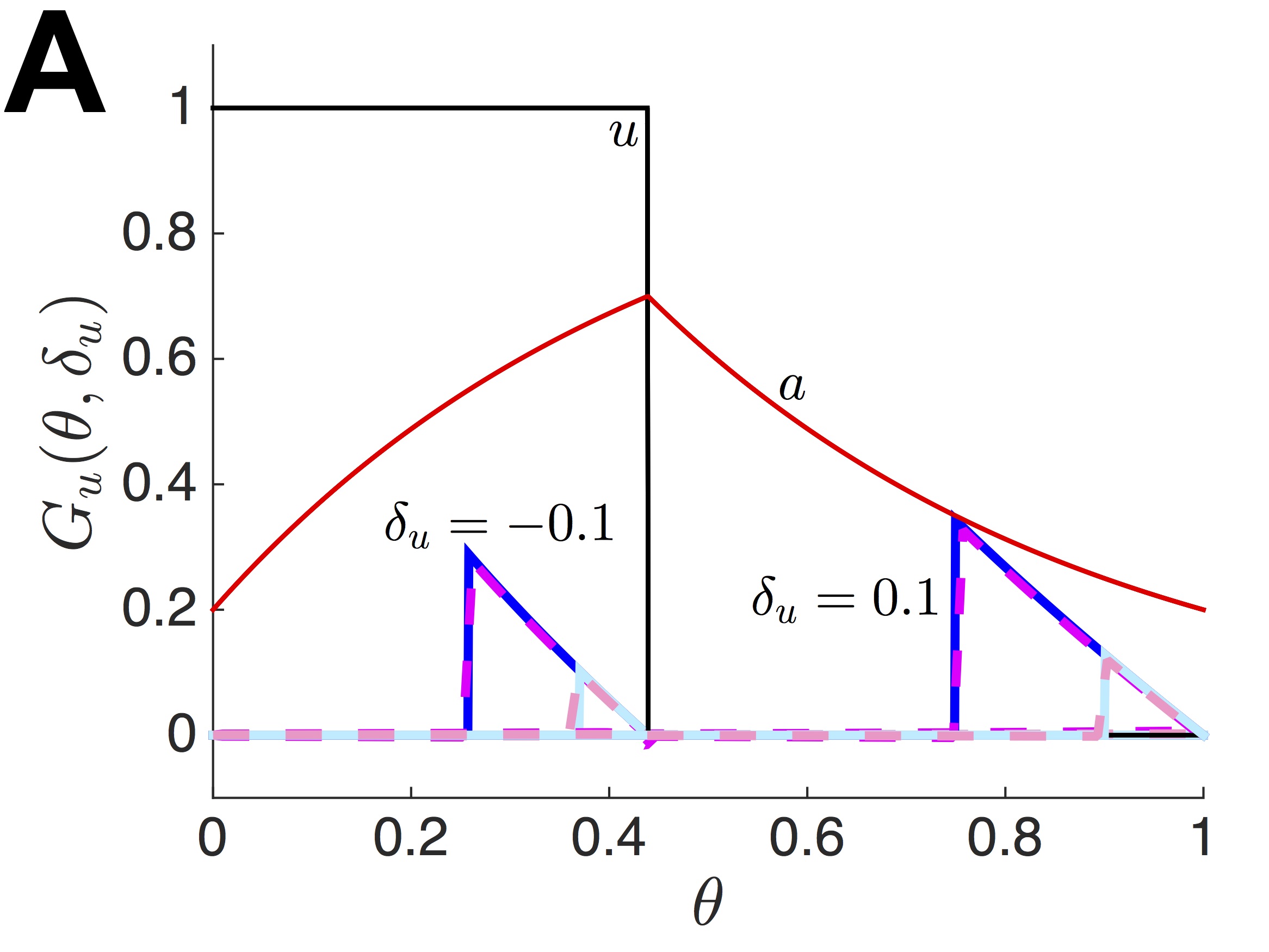} \includegraphics[width=4cm]{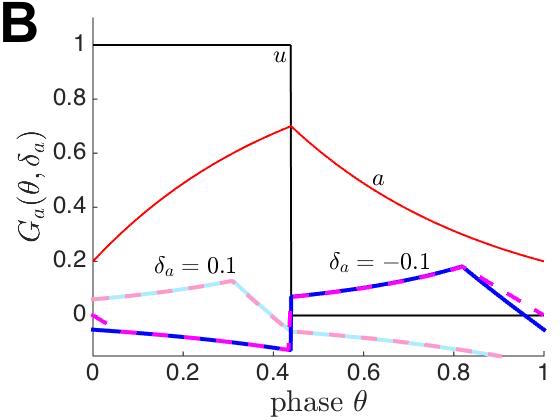} \includegraphics[width=4cm]{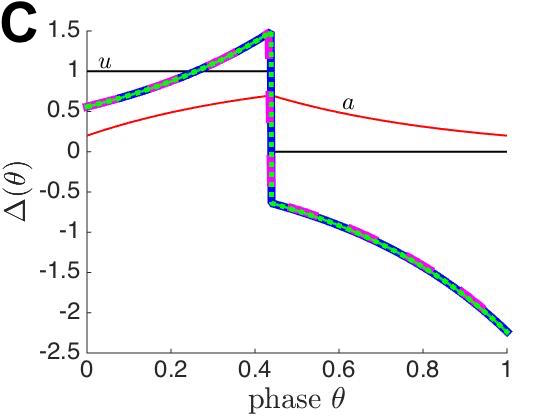} \end{center}
\caption{Phase response curves of the fast-slow timescale separated system $\tau \gg 1$. ({\bf A},{\bf B}) Amplitude $\delta_u$- and $\delta_a$-dependent phase response curves $G_u(\theta, \delta_u)$ and $G_a(\theta, \delta_u)$ characterizing phase advances/delays resulting from perturbation of neural activity $u$ and adaptation $a$. We compare analytical formulae (solid lines) to numerically computed PRCs (dashed lines). ({\bf C}) Phase response curve associated with perturbations of the adaptation variable $a$ in the small amplitude $0 < |\delta_a| \ll 1$ limit. We compare the large amplitude formula (solid line) determined by (\ref{ptcfora}) to the linear approximation (dotted line) given by (\ref{prcalin}) to numerical computations (dashed line).}
\label{fastslowprc}
\end{figure}

Here, we decompose the impact of perturbations to the $u$ and $a$ variables. We begin by studying the impact of perturbations $\delta_u$ to the activity variable $u$. We can directly compute
\begin{align*}
\tilde{u}_0(t) = H(I + \alpha \left[ u_0(t) + \delta_u \right] - a_0(t)).
\end{align*}
Thus, the singular system (\ref{fastsub}) will be unaffected by such perturbations if $\sgn (I + \alpha [u + \delta_u] - a  ) = \sgn (I + \alpha u - a  )$. This is related to the flatness of the susceptibility function $Z_u$ over much of the time domain in Fig. \ref{prcsig}{\bf D},{\bf E},{\bf F}. However, if $\sgn (I + \alpha [u + \delta_u] - a  ) \neq \sgn (I + \alpha u - a  )$, then $\tilde{u}_0(t) = 1-u_0(t)$, as detailed in the following piecewise smooth map:
\begin{align*}
\begin{array}{ll} 
u_0(t) = 0 \mapsto \tilde{u}_0(t) = 1 \ & : \delta_u > - (I - a_0(t))/ \alpha > 0 , \\[1ex]
u_0(t) = 0 \mapsto \tilde{u}_0(t) = 0 \ & : \delta_u < - (I - a_0(t))/ \alpha > 0 , \\[1ex]
u_0(t) = 1 \mapsto \tilde{u}_0(t) = 0 \ & : - \delta_u <  - ( I + \alpha - a_0(t))/ \alpha < 0, \\[1ex]
u_0(t) = 1 \mapsto \tilde{u}_0(t) = 1 \ & : - \delta_u >  - ( I + \alpha - a_0(t))/ \alpha < 0, \\
\end{array}
\end{align*}
where $(u_0(t),a_0(t))$ are defined by (\ref{Hpersol}). The formula (\ref{invertime}) can then be utilized to compute the updated relative time $\tilde{t}_0 := t_0(\tilde{u}_0,\tilde{a}_0)$, finding
\begin{align} \tilde{t}_0 = \left\{
\begin{array}{ll} 
\tau \ln \left[ (\phi - I)(I+ \alpha)/a_0/(\phi - \alpha - I) \right] \ & : \delta_u >  ( I + \alpha - a_0)/ \alpha > 0, \ u_0 = 1 \\[1ex]
T + \tau \ln \left[ ( \phi - I)/(\phi - a_0) \right] \ & : -\delta_u > (a_0 - I)/ \alpha > 0 , \ u_0 = 0 \\[1ex]
t_0(u_0,a_0) \ & : {\rm otherwise},
\end{array} \right.  \label{ptcforu}
\end{align} 
where $a_0 = \phi - (\phi - I)\e^{-t_0/\tau}$ if $u_0=1$ and $a_0 = (I+\alpha)\e^{-(t_0-T_1)/\tau}$ if $u_0 = 0$. We can refer to the function $\tilde{t}_0/T$, where $\tilde{t}_0$ is defined by (\ref{ptcforu}), as the {\em phase transition curve} for $u$ perturbations. Thus, the function $G_u ( \theta, \delta_u) = (\tilde{t}_0 -t_0)/T$ will be the {\em phase response curve}, where $\theta = t_0/T$, and phase advances occur for positive values and phase delays occur for negative values. We plot the function $G_u(\theta, \delta_u)$ in Fig. \ref{fastslowprc}{\bf A} for different values of $\delta_u$, demonstrating the nontrivial dependence on the perturbation amplitude is not simply a rescaling but an expansion of the non-zero phase shift region. Due to the singular nature of the fast-slow limit cycle (\ref{Hpersol}), the size of the phase perturbation has a piecewise constant dependence on the amplitude of the $u$ perturbation. Note, this formulation allows us to quantify phase shifts that would not be captured by a perturbative theory for phase sensitivity functions, as computed for the general system in (\ref{adjeqns}).

For perturbations $\delta_a$ of the adaptation variable $a$, there is a more graded dependence of the phase advance/delay amplitude on the perturbation amplitude $\delta_a$. We expect this, as it was a property we observed in $Z_a$ as we varied parameters in Fig. \ref{prcsig}. We can partition the limit cycle $(u_0(t),a_0(t))$ into four different regions: two advance/delay regions of exponential saturation and two early threshold crossings. First, note if $u_0(t) = 1$ and $a_0(t) + \delta_a < I + \alpha$, then
\begin{align}
\tilde{u}_0(t) = 1, \hspace{5mm} \tilde{a}_0(t) = \phi - (\phi - I)\e^{-t/\tau} + \delta_a,  \label{da1}
\end{align}
so $\tilde{t}_0 = T_1 - t_w$ with $t_w = \tau \ln \left[ (\phi - a_0 - \delta_a)/(\phi - I - \alpha) \right]$, but if $a_0(t) + \delta_a > I + \alpha$, then
\begin{align}
\tilde{u}_0(t) = 0, \hspace{5mm} \tilde{a}_0(t) = \phi - (\phi - I)\e^{-t/\tau} + \delta_a. \label{da2}
\end{align}
Determining the relative time of the perturbed variables $(\tilde{u}_0(t), \tilde{a}_0(t))$ in (\ref{da1}) is straightforward using the mapping (\ref{invertime}). However, to determine the relative time described by (\ref{da2}), we compute the time, after the perturbation, until $\tilde{a}_0 (t) = I+\alpha$, which will be $t_w = \tau \ln \left[ (a_0 + \delta_a)/(I+\alpha) \right]$, so $\tilde{t}_0 = T_1 - t_w$.  Second, note if $u_0(t) = 0$ and $a_0(t) + \delta_a > I $, then
\begin{align}
\tilde{u}_0(t) = 0, \hspace{5mm} \tilde{a}_0(t) = (I+ \alpha) \e^{-(t-T_1)/\tau} + \delta_a,  \label{da3}
\end{align}
so $\tilde{t}_0 = T - t_w$ with $t_w = \tau \ln \left[ (a_0 + \delta_0)/I \right]$, but if $a_0(t) + \delta_a < I$, so that it is necessary that $\delta_a<0$, then
\begin{align}
\tilde{u}_0(t) = 1, \hspace{5mm} \tilde{a}_0(t) = (I+\alpha) \e^{-(t-T_1)/\tau} + \delta_a.  \label{da4}
\end{align}
In the case of (\ref{da4}), we note that $t_w = \tau \ln \left[ (\phi - a_0 - \delta_a)/(\phi - I) \right]$, so $\tilde{t}_0 = T - t_w$. Combining our results, we find we can map the relative time to the perturbed relative time as
\begin{align}
\tilde{t}_0  = \left\{
\begin{array}{ll} 
T_1 - \tau \ln \left[ (\phi - a_0 - \delta_a)/(\phi - I - \alpha ) \right] \ & : \delta_a < I + \alpha - a_0, \ u_0 = 1, \\[1ex]
T_1 - \tau \ln \left[ (a_0 + \delta_a)/(I + \alpha ) \right] \ & :  \delta_a > I + \alpha - a_0, \ u_0 = 1,  \\[1ex]
T - \tau \ln \left[ (a_0 + \delta_a)/I \right] \ & :  \delta_a > I  - a_0, \ u_0 = 0,  \\[1ex]
T - \tau \ln \left[ (\phi - a_0 - \delta_a )/ (\phi - I) \right] \ & :  \delta_a < I  - a_0, \ u_0 = 0,
\end{array} \right.  \label{ptcfora}
\end{align}
where $a_0 = \phi - ( \phi - I) \e^{-t_0/\tau}$ if $u_0 = 1$ and $a_0 = (I+ \alpha) \e^{- (t_0 - T_1)/\tau}$ if $u_0 = 0$. Again, we have a {\em phase transition curve} given by the function $\tilde{t}_0/T$ and {\em phase response curve} given by $G_a(\theta, \delta_a ) = (\tilde{t}_0 - t_0)/T$, where $\theta = t_0/T$. As opposed to the case of $u$ perturbations, the phase perturbation here depends smoothly on the amplitude of the $a$ perturbation $\delta_a$.

Furthermore, we can obtain a perturbative description of the phase response curve for the singular system (\ref{fastsub}) in two ways: (a) Taylor expand the amplitude-dependent phase response curve expressions defined by (\ref{ptcforu}) and (\ref{ptcfora}) and truncate to linear order or (b) solving the adjoint equation (\ref{adjeqns}) in the case of a Heaviside firing rate (\ref{H}) and long adaptation timescale $\tau \gg 1$. We begin with the first derivation, which simply requires differentiating (\ref{ptcforu}) to demonstrate that the {\em infinitesimal phase response curve (iPRC)} associated with perturbations of the $u$ variable is zero almost everywhere. However, differentiating (\ref{ptcfora}) reveals that the iPRC associated with perturbations of the adaptation variable $a$ is given by the piecewise smooth function
\begin{align}  \label{prcalin}
Z_a (t) = \left\{ \begin{array}{ll} \D \frac{\tau}{T(\phi - I)s} \e^{t/\tau} \ & : t \in (0,  T_1), \\[1ex]
\D - \frac{\tau}{T(I+\alpha)} \e^{(t - T_1)/\tau} \ & : t \in (T_1, T).
\end{array} \right.
\end{align}
Furthermore, note we could derive the same result by solving the adjoint equations (\ref{adjeqns}) in the case of Heaviside firing rate (\ref{H}), so that
\begin{subequations}   \label{Hadjeq}
\begin{align}
\dot{Z}_u(t) &= - Z_u(t) + \alpha \delta ( \alpha u_0(t) - a_0(t) + I) Z_u(t) + \phi Z_a(t) / \tau, \\
\dot{Z}_a(t) &= - \delta( \alpha u_0(t) - a_0 (t) + I) Z_u(t) + Z_a (t) / \tau.
\end{align}
\end{subequations}
Note, we have reversed time $t \mapsto - t$, so we can simply solve the system forward. Furthermore, we can use the identity
\begin{align}
\delta( \alpha u_0(t) - a_0(t) + I) = \frac{\delta(t)}{u'(0) - a'(0)} + \frac{\delta(t-T_1)}{u'(T_1) - a'(T_1)}.
\end{align}
Utilizing the separation of timescales, $\tau \gg 1$, we find that almost everywhere (except where $t=0, T_1, T$), we have that (\ref{Hadjeq}) becomes the system
\begin{align}
\dot{Z}_u(t) = -Z_u(t), \hspace{6mm} \tau \dot{Z}_a(t) = Z_a(t).\label{Hadjeq2}
\end{align}
As before $Z_u(t)$ will be zero almost everywhere, whereas $Z_a(t) = A(t) \e^{t/\tau}$, where $A(t)$ is a piecewise constant function taking two different values on $t \in (0,T_1)$ and $t \in (T_1, T)$, determined by considering the $\delta$ distribution terms. This indicates how one would derive the formula (\ref{prcalin}) using the adjoint equations (\ref{Hadjeq}). 

Note, in previous work \citep{jayasuriya12}, we explored the entrainment of slowly adapting populations to external forcing, comprised of smooth and non-smooth inputs to the system (\ref{single}). In the next section, we explore the impact of external noise forcing on the slow oscillations of (\ref{single}), subsequently demonstrating that noise can be utilized to entrain the up and down states of two distinct networks.

\section{Impact of noise on the timing of up/down states}
\label{1dnoise}

We now study the effects of noise on the duration of up and down states of the single population model (\ref{single}). Switches between high and low firing rate states can occur at irregular intervals \citep{sanchezvives00}, suggesting internal or external sources of noise determine state changes. This section focuses on how noise can reshape the mean duration of up and down residence times. Due to our findings in the previous sections, we focus on noise applied to the adaptation variable in this section. As we have shown, very weak perturbations to the neural activity variable have a negligible effect on the phase of oscillations. Analytic calculations are presented for the piecewise smooth system with Heaviside firing rate (\ref{H}), as accurate approximations of the mean up and down state durations can be computed. 

\begin{figure}
\begin{center} \includegraphics[width=5.5cm]{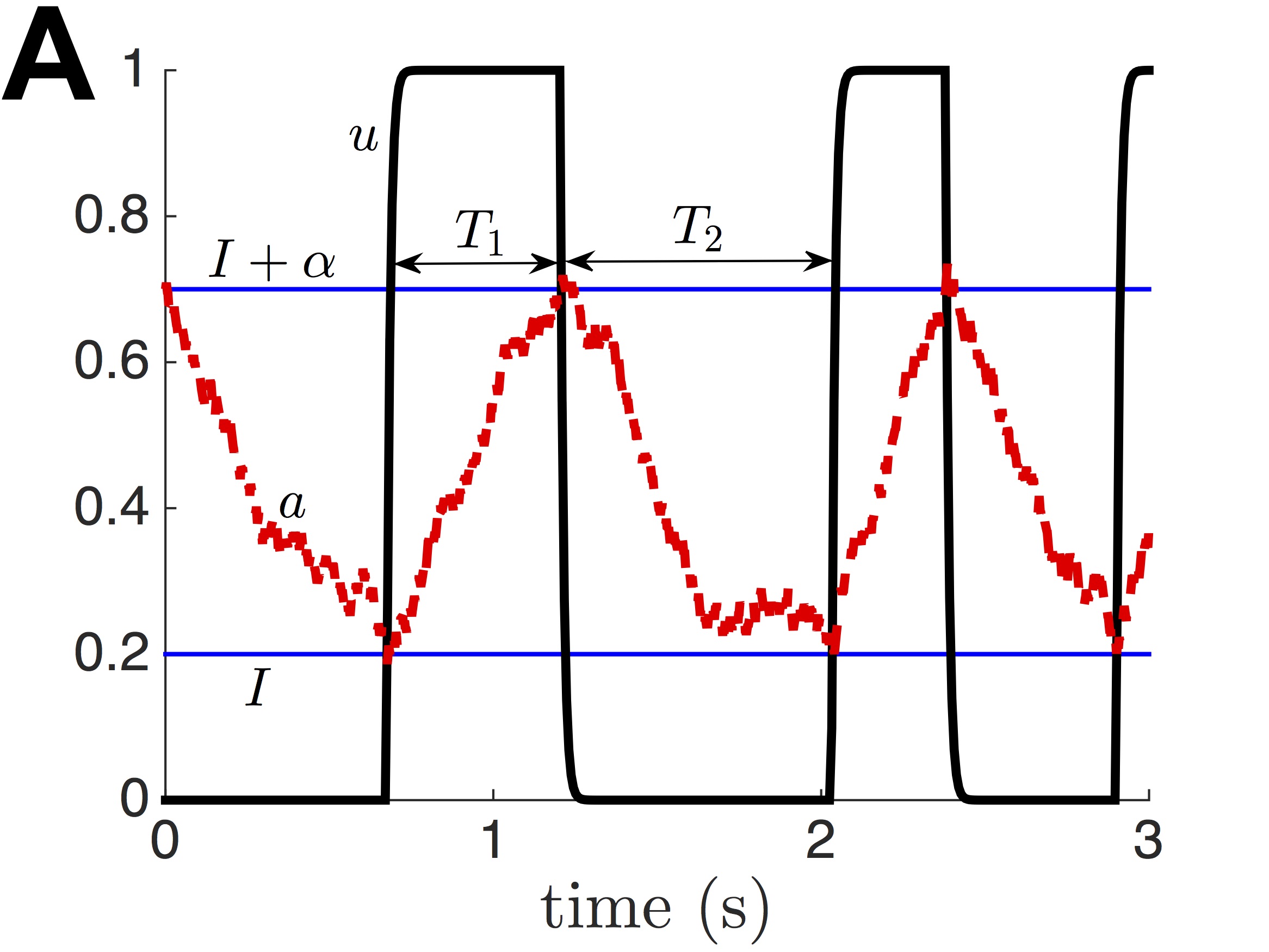} \includegraphics[width=5.5cm]{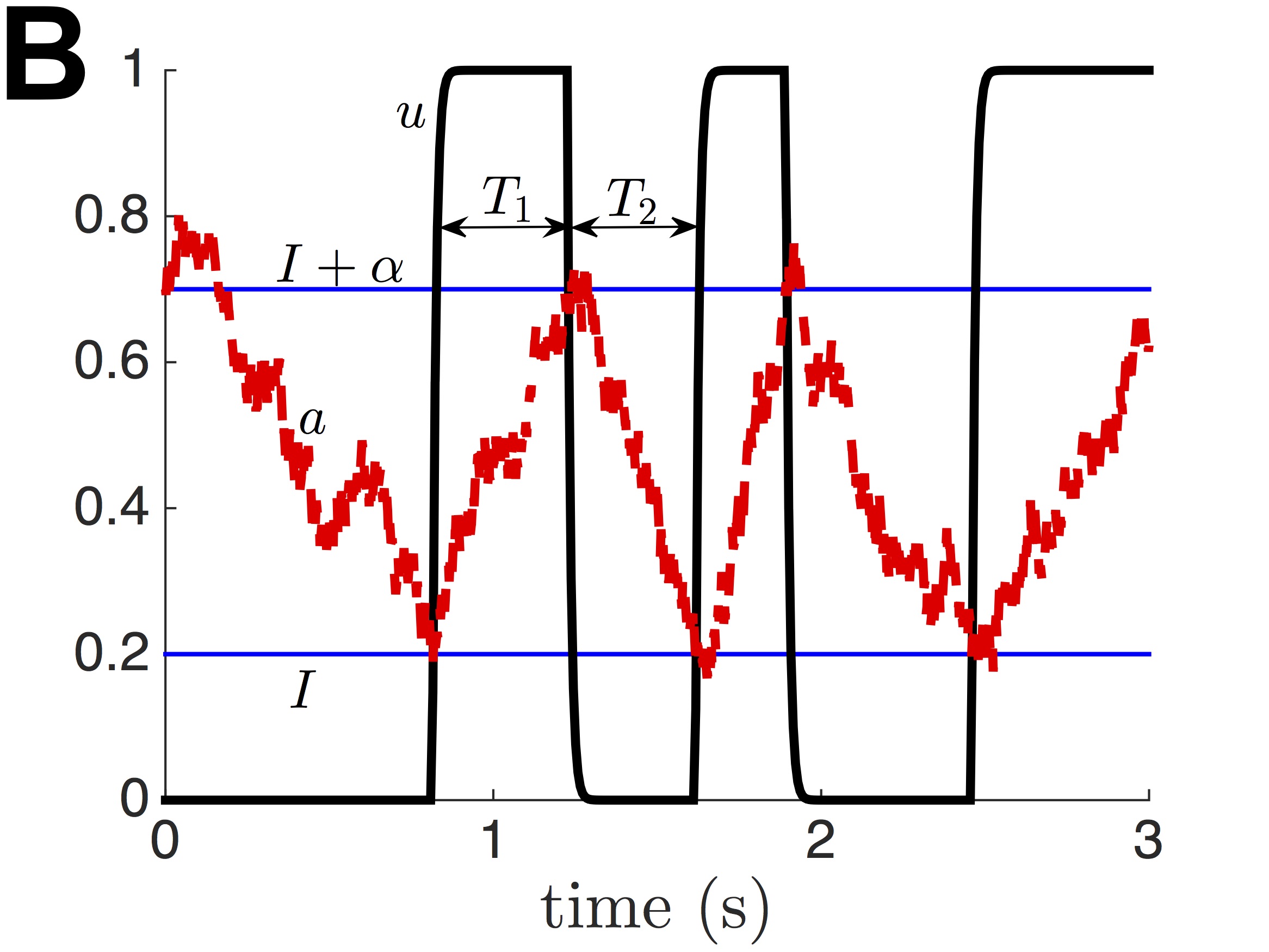} \\ \includegraphics[width=5.5cm]{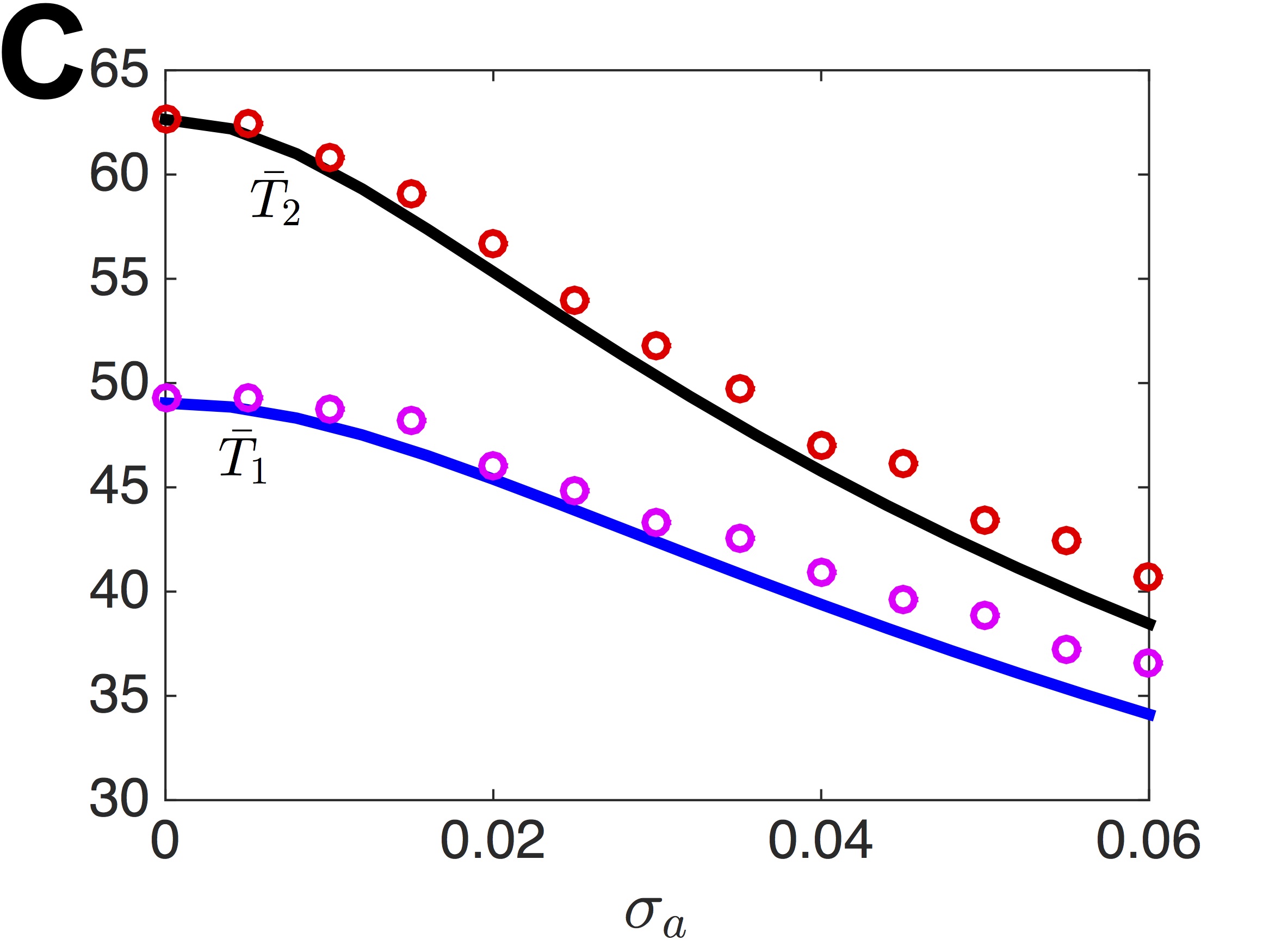}  \includegraphics[width=5.5cm]{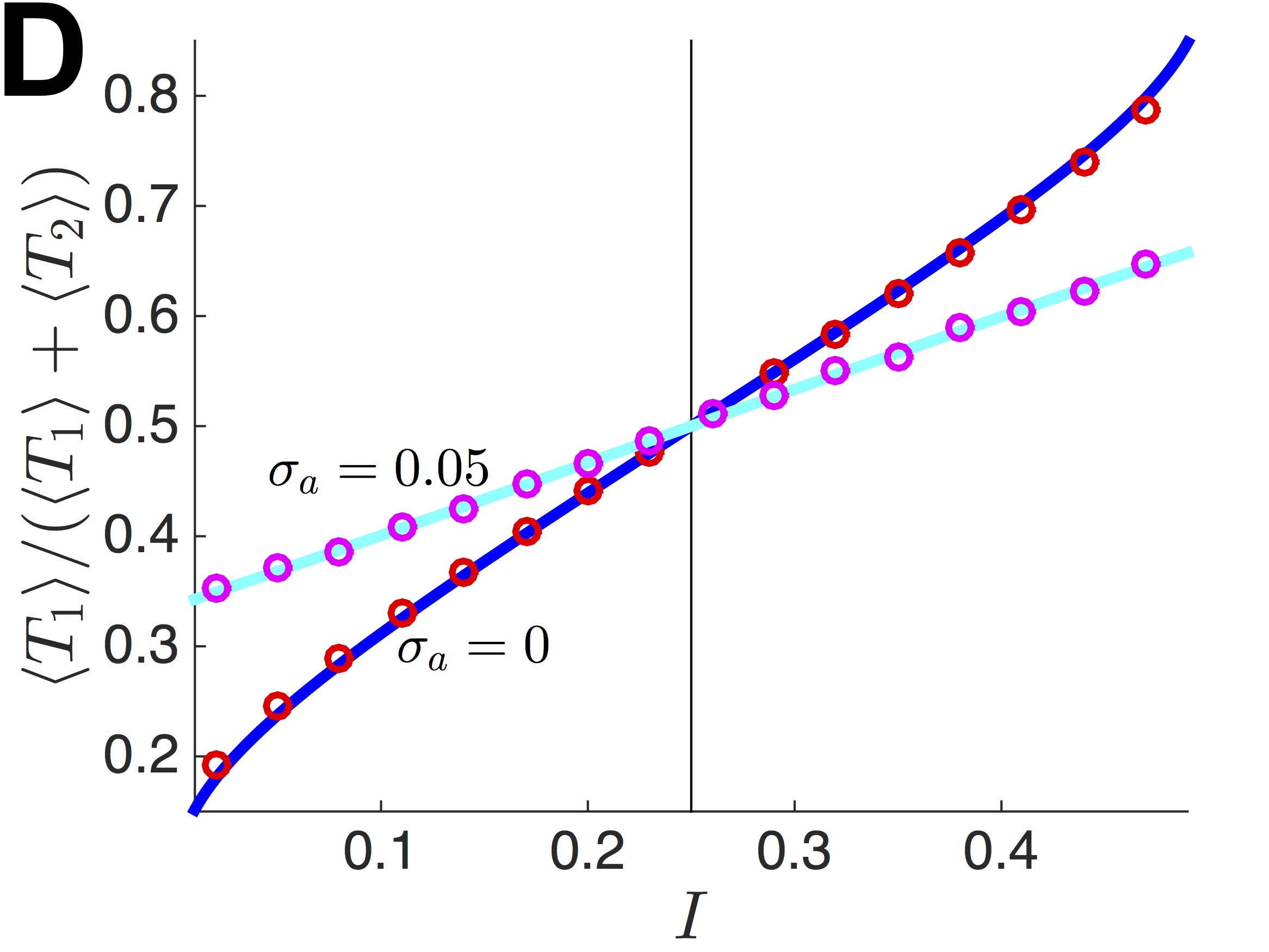} \end{center}
\caption{Noise alters the duration of up and down states. ({\bf A}) Numerical simulation of the stochastically driven population model (\ref{stochmod}) demonstrates up and down state durations (e.g., $T_1$ and $T_2$) are variable when driven by adaption noise $\xi_a$ with $\langle \xi_a^2 \rangle = \sigma_a^2 t$, $\sigma_a = 0.01$. Switches are determined by the threshold crossings of the adaptation variable $a(t)=I$ and $a(t) = I+ \alpha$. ({\bf B}) Up/down states become more variable when the noise amplitude $\sigma_a = 0.02$. ({\bf C}) Mean durations of the up and down state, $\langle T_1 \rangle$ and $\langle T_2 \rangle$, decrease as a function noise amplitude $\sigma_a$. ({\bf D}) Impact of noise $\sigma_a$ on the balance of up to down state durations $\bar{T}_1/\bar{T}_2$ as input $I$ is varied. Firing rate is given by the Heaviside function (\ref{H}). Other parameters are $\alpha = 0.5$, $\phi = 1$, and $\tau = 50$. }
\label{snoiseper}
\end{figure}

Our approach is to derive expressions for the mean first passage times of both the up and down state ($\bar{T}_1$ and $\bar{T}_2$) of the stochastic population model (\ref{stochmod}). Focusing on adaptation noise allows us to utilize the separation of fast-slow timescales, and recast the pair of equations as a stochastic-hybrid system
\begin{align*}
u(t) &= H( (\alpha u(t) + I - a(t)), \\
\d {a}(t) &= \left[ -a(t) + \phi u(t) \right] \d t/ \tau + \d \xi_a(t),
\end{align*}
where $\xi_a$ is white noise with mean $\langle \xi_a \rangle = 0$ and variance $\langle \xi_a^2 \rangle = \sigma_a^2 t$. To begin, assume the system has just switched to the up state, so the initial conditions are $u(0) = 1$ and $a(0) = I$. Determining the amount of time until a switch to the down state requires we calculate the time $T_1$ until the threshold crossing $a(T_1) = I + \alpha$ where $a(t)$ is determined by the stochastic differential equation (SDE)
\begin{align*}
\d a(t) = \left[ - a(t) + \phi \right] \d t/ \tau + \d \xi_a,
\end{align*}
which is the well-known threshold crossing problem for an Ornstein-Uhlenbeck process \cite{gardiner04}. The mean $\bar{T}_1$ of the passage time distribution is thus given by defining the potential $V(a)=\frac{a^2}{2 \tau} - \frac{\phi a}{\tau}$ and computing the integral
\begin{align*}
\bar{T}_1 &= \frac{1}{\sigma_a^2} \int_{I}^{I + \alpha} \int_{- \infty}^{x} \e^{\left[ V(x) - V(y) \right]/ \sigma_a^2} \d y \d x.
\end{align*}
Next, note that the duration of the down state $T_2$ will be the amount of time until the threshold crossing $a(T_2) = I$ given $u(0) = 0$ and $a(0) = I+ \alpha$, where $a(t)$ obeys the SDE
\begin{align*}
\d a(t) = \left[ -a(t) \right] \d t/ \tau + \d \xi_a(t).
\end{align*}
Again, defining the potential $V(a) = \frac{a^2}{2 \tau}$, we can compute
\begin{align*}
\bar{T}_2 &= \frac{1}{\sigma_a^2} \int_{-I-\alpha}^{-I} \int_{- \infty}^{x} \e^{\left[ V(x) - V(y) \right]/ \sigma_a^2} \d y \d x.
\end{align*}
We compare the theory we have developed utilizing passage time problems to residence times computed numerically in Fig. \ref{snoiseper}{\bf C}. Notice that increasing the noise amplitude tends to shorten both up and down state durations on average, due to early threshold crossings of the variable $a(t)$.

Furthermore, we can examine how noise reshapes the relative balance of up versus down state durations. Specifically, we will explore how the relative fraction of time the up state persists $\bar{T}_1/(\bar{T}_1 + \bar{T}_2)$ changes with noise intensity $\sigma_a$ and input $I$. First, notice that, in the absence of noise the ratio
\begin{align}
\frac{T_1}{T_1+T_2} = \frac{\ln \left[ (\phi - I)/(\phi - \alpha - I) \right]}{\ln \left[ (I+ \alpha)(\phi - I)/I(\phi - \alpha - I) \right]}.   \label{updownrat}
\end{align}
The up and down state have equal duration when $T_1/(T_1 + T_2) = 1/2$, or when the input $I = (\phi - \alpha)/2$, as shown in Fig. \ref{snoiseper}{\bf D}. Interestingly, this is the precise input value at which the period obtains a minimum, as we demonstrated in section \ref{periodicsoln}. Along with our plot of (\ref{updownrat}) in the noise-free case ($\sigma_a = 0$), we also study the impact of noise on this measure of up-down state balance. Noise leads to up and down state durations becoming more similar, so the ratio (\ref{updownrat}) of the means $\bar{T}_1$ and $\bar{T}_2$ flattens as a function of the input $I$. This is due to the fact that long durations, wherein the variable $a(t)$ occupies the tail of exponentially saturating functions $A_0 + A_1 \e^{-t/\tau}$, are shortened by early threshold crossings due to the external noise forcing.

\section{Synchronizing two uncoupled populations}
\label{ssynch}

Now we demonstrate that common noise can synchronize the up and down states of two distinct and uncoupled populations. We begin with the case of identical noise and then, in section \ref{indepnos}, relax these assumptions to show that some level of coherence is still possible when each population has an intrinsic and independent source of noise. This is motivated by the observation that the neural Langevin equation derived in the large system-size limit of a neural master equation tends to possess intrinsic noise in each population, in addition to an extrinsic common noise term \citep{bressloff11}. As we will show, intrinsic noise tends to disrupt the phase synchronization due to extrinsic noise.

\begin{figure}
\begin{center} \includegraphics[width=5.5cm]{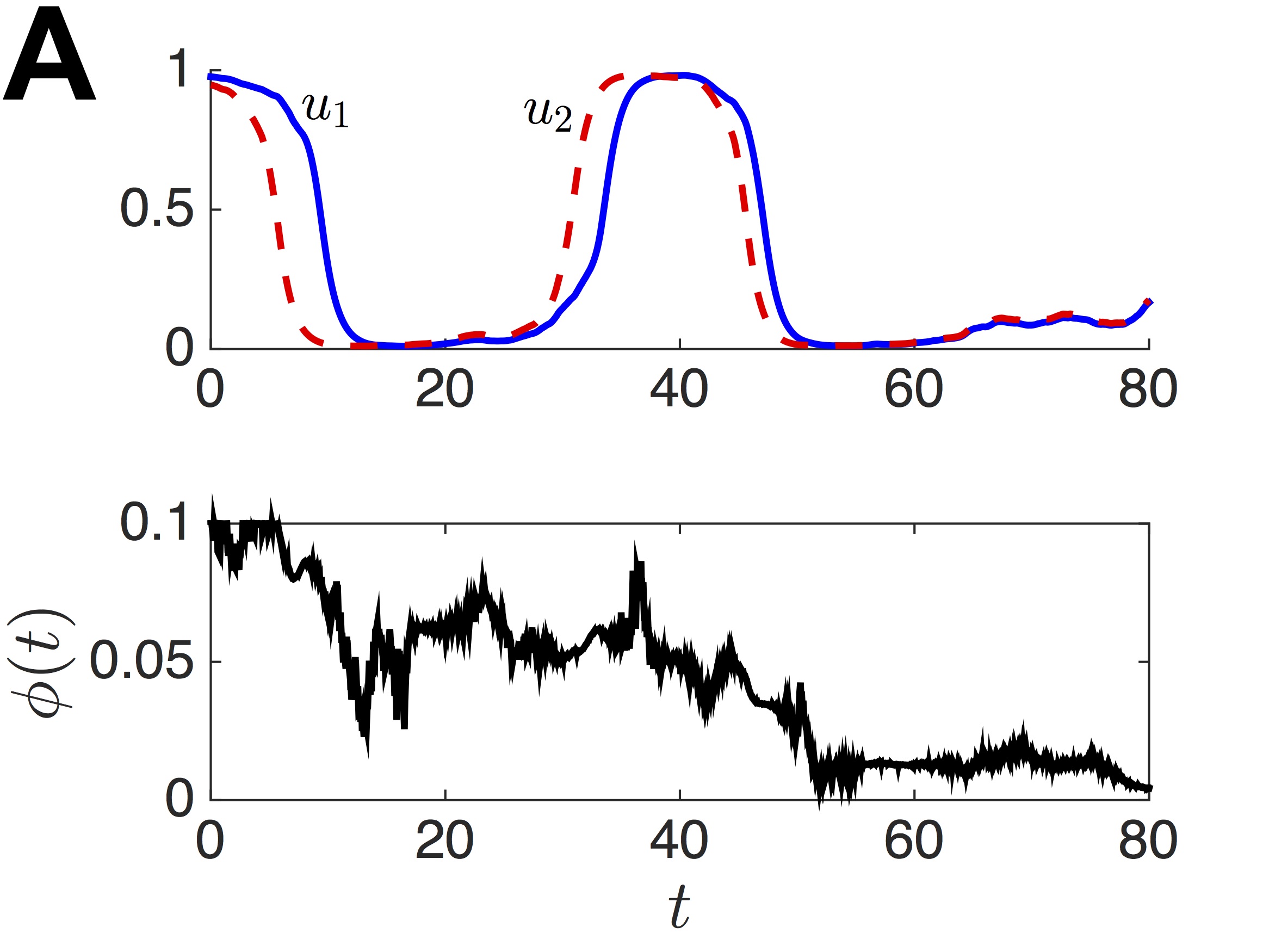} \includegraphics[width=5.5cm]{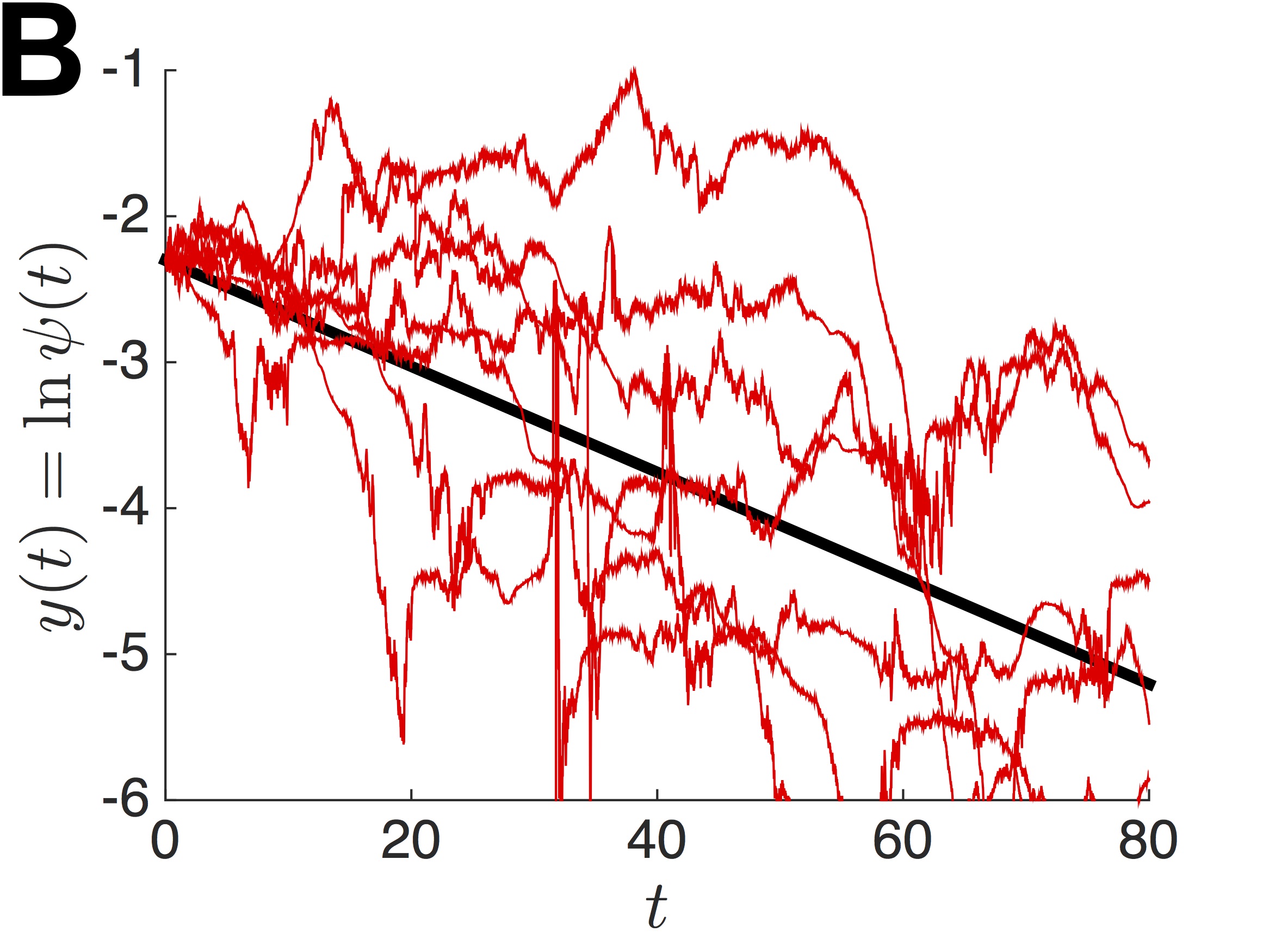} \\ \includegraphics[width=5.5cm]{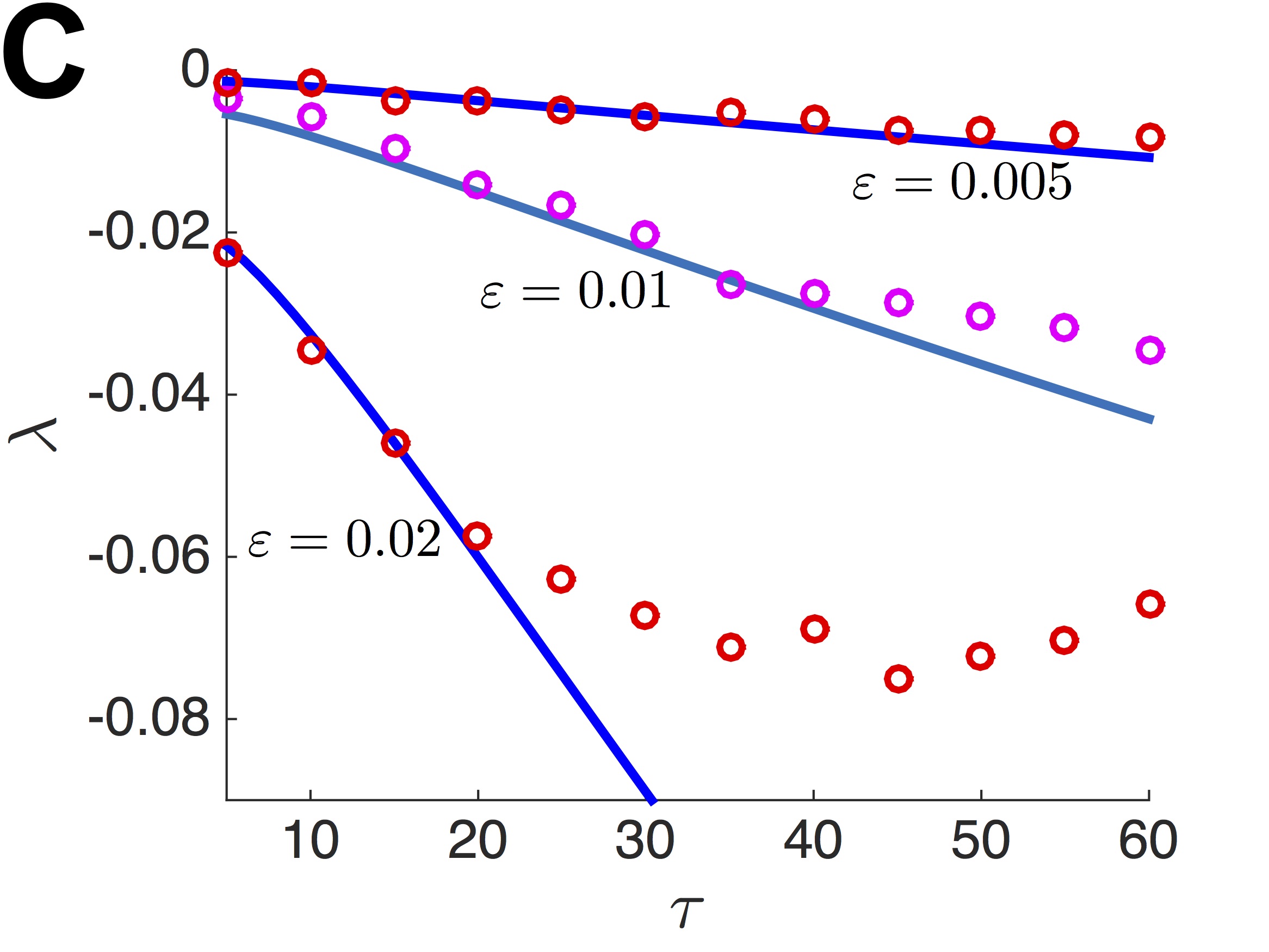} \includegraphics[width=5.5cm]{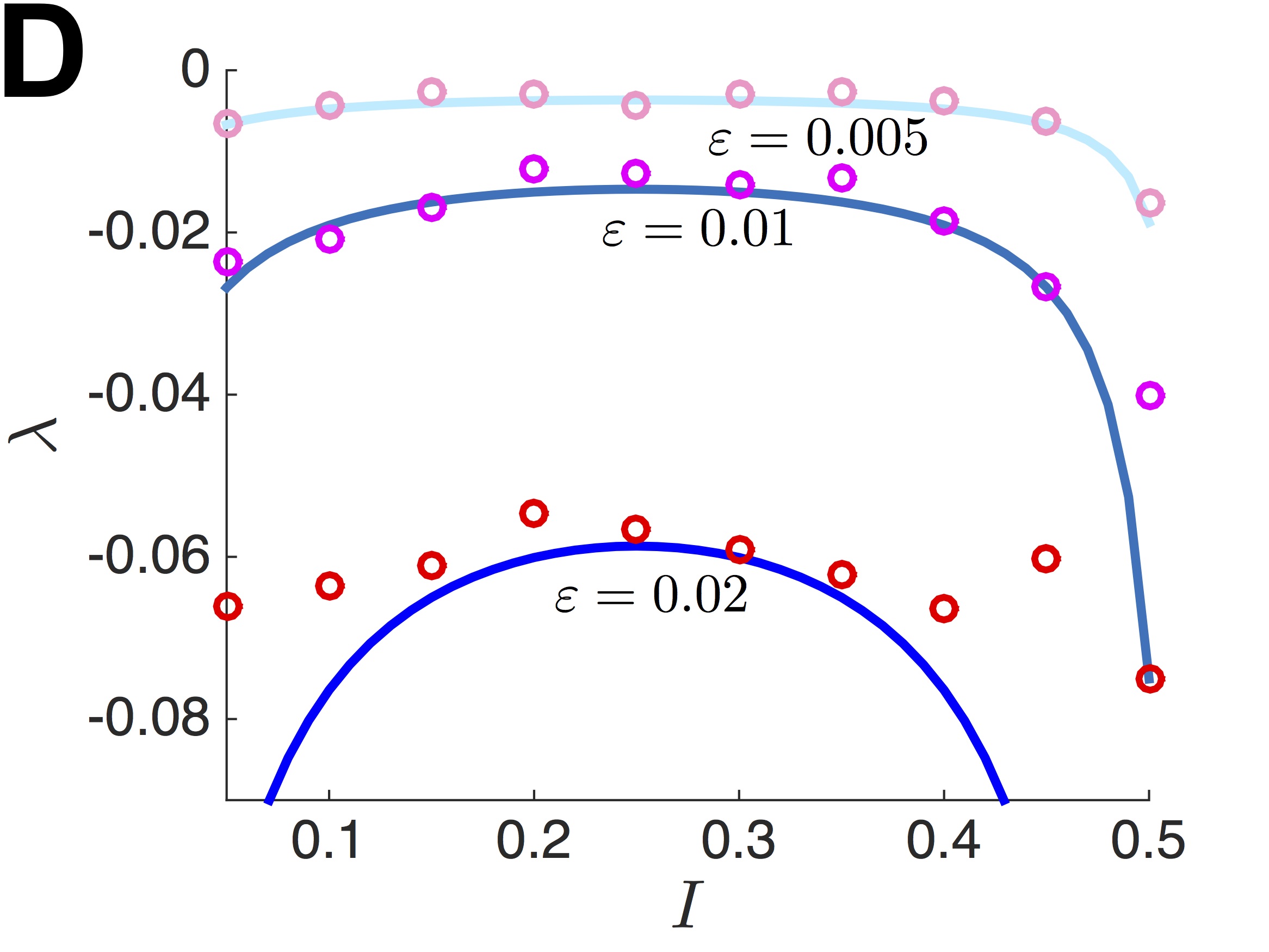} \end{center}
\caption{Synchronizing slow oscillations in two uncoupled populations described by (\ref{dual}) with sigmoidal firing rate (\ref{sig}). ({\bf A}) Single realization of the system (\ref{dual}) driven by common noise $\xi_a$ to the adaptation variable ($\langle \xi_a^2 \rangle = \ve^2 t$, $\ve = 0.01$) with input $I=0.2$ and adaptation timescale $\tau = 50$. Notice that the phase difference $\psi (t) = \Delta_1(t) - \Delta_2(t)$ roughly decreases over time. ({\bf B}) Plot of the log of the phase difference $y(t) = \ln \psi (t)$ for several realizations (thin lines) compared with the theory (thick line) of the mean $y(0) + \lambda t$ computed using the Lyapunov exponent (\ref{lyapapprox}). ({\bf C}) Lyapunov exponent $\lambda$ decreases as a function of the adaptation timescale $\tau$, for $I = 0.2$. We compare numerical simulations (dots) to theory (solid). ({\bf D}) Lyapunov exponent $\lambda$ varies non-monotonically with the strength of the input $I$. Other parameters are $\alpha = 0.5$, $\gamma = 15$, and $\phi = 1$.}
\label{figstochsynch}
\end{figure}

To begin, we recast the stochastic system (\ref{dual}), describing a pair of adapting noise-driven neural populations, as a pair of phase equations:
\begin{subequations} \label{dstrat}
\begin{align}
\d \theta_1 (t) &= \omega \d t + \Z( \theta_1(t) ) \cdot \d \bxi (t), \\
\d \theta_2 (t) &= \omega \d t + \Z( \theta_2(t)) \cdot \d \bxi (t), 
\end{align}
\end{subequations}
where $\theta_1$ and $\theta_2$ are the phase of the first and second neural populations. As we demonstrate in Fig. \ref{figstochsynch}{\bf A}, this introduction of common noise tends to drive the oscillation phases $\theta_1(t)$ and $\theta_2(t)$ toward one another. Note that since the governing equations of both populations are the same, then the phase sensitivity function $\Z (\theta)$ will be the same for both. Furthermore, the synchronized solution $\theta_1(t) = \theta_2(t)$ is absorbing -- once the phases synchronize, they remain so. We can analytically calculate the Lyapunov exponent $\lambda$ of the synchronized state to determine its stability. In particular, we will be interested in how this stability depends on the parameters that shape the dynamics of adaptation.

We next convert the pair of Stratonovich differential equations into a equivalent pair of Ito differential equations:
\begin{subequations} \label{dito}
\begin{align}
\d \theta_1 (t) &= \left[ \omega  + \Z'(\theta_1(t))^T \Db \Z (\theta_1(t)) \right] \d t + \Z (\theta_1 (t)) \cdot \d \bxi (t),  \\
\d \theta_2 (t) &= \left[ \omega + \Z'(\theta_2 (t))^T \Db \Z (\theta_2(t)) \right] \d t + \Z ( \theta_2 (t)) \cdot \d \bxi (t),
\end{align}
\end{subequations}
introducing a drift term due to our change in definition of the noise term. Now, to determine stability of the synchronized state $\theta_1 (t) = \theta_2(t)$, we assume we are infinitesimally close to it. We define $\psi (t) = \theta_1(t) - \theta_2(t)$ and require $|\psi (t) | \ll 1$. Linearizing the system of Ito differential equations (\ref{dito}) with respect to $\psi (t)$ then yields
\begin{align}
\d \psi (t) = \psi (t) \left[ \left( \Z' (\theta(t) )^T \Db \Z ( \theta(t) ) \right)' \d t + \Z'(\theta(t)) \cdot \d \bxi \right], \label{pdiff}
\end{align}
where $\theta (t)$ obeys either one of the equations in (\ref{dito}). Applying the change of variables $y (t) = \ln \psi (t) $, we can rewrite (\ref{pdiff}) as
\begin{align}
\d y (t) = \left( \Z' \Db \Z \right)' \d t - \left( \Z'^T \Db \Z' \right) \d t + \Z' \cdot \d \bxi (t).  \label{ylog}
\end{align}
Notice, on average, the log of the phase difference $y(t)$ tends to decrease over time (Fig. \ref{figstochsynch}{\bf B}), indicating the phases $\theta_1$ and $\theta_2$ move toward one another. Subsequently, we can integrate equation (\ref{ylog}) to determine the mean drift of $y(t)$
\begin{align*}
\lambda := \lim_{t \to \infty} \int_0^t \left[ \left( \Z'(\theta(s)) \Db \Z(\theta(s)) \right)' - \left( \Z'^T(\theta(s)) \Db \Z'(\theta(s)) \right)  \right] \d s,
\end{align*}
so the phase difference $\psi (t)$ will tend to decay grow if the Lyapunov exponent $\lambda <0$ ($\lambda >0$), and the synchronous state will be stable (unstable). Now, utilizing ergodicity of (\ref{ylog}), we can compute $\lambda$ using the ensemble average across realizations of $\Z' (\theta(t)) \cdot \bxi (t)$, so
\begin{align}
\lambda &= \int_0^1 P_s (\theta) \left[ \left( \Z'(\theta) \Db \Z(\theta) \right)' - \left( \Z'^T(\theta) \Db \Z'(\theta) \right)  \right] \d \theta,  \label{ensavg}
\end{align}
where $P_s(\theta)$ is the steady state distribution of $\theta$. Since noise is weak ($\Db_{jk} \ll 1$, $j,k=1,2$), we can approximate the distribution as constant $P_s(\theta) = 1$. Upon applying this to the integrand of (\ref{ensavg}) and noting the periodicity of $\Z (\theta)$, we find we can approximate the Lyapunov exponent
\begin{align}
\lambda &= - \int_0^1 \Z'^T(\theta) \Db \Z'(\theta) \d \theta. \label{lyapapprox}
\end{align}
Assuming noise to the activity variable $u$ and adaptation variable $a$ is not correlated, $\Db$ will be diagonal. In this case, we can further decompose the phase sensitivity function into its Fourier expansion
\begin{align*}
\Z (\theta) = \sum_{k=0}^{\infty} \ab_k \sin (2 \pi k \theta) + \bb_k \cos (2 \pi k \theta),
\end{align*}
where $\ab_k = (\ab_{k1}, \ab_{k2})^T$ and  $\bb_k = (\bb_{k1}, \bb_{k2})^T$ are vectors in $\R^2$ so that
\begin{align*}
\Z' (\theta) = \sum_{k=0}^{\infty} 2 \pi k \left[  \ab_k \cos (2 \pi k \theta) - \bb_k \sin (2 \pi k \theta) \right],
\end{align*}
and we can expand the terms in (\ref{lyapapprox}) to yield
\begin{align*}
\lambda  = - \sum_{k = 0}^{\infty} 2 \pi^2 k^2 \left[ \left( \ab_{k1}^2 + \bb_{k1}^2 \right) D_{11} +  \left( \ab_{k2}^2 + \bb_{k2}^2 \right) D_{22} \right].
\end{align*}
Thus, as long as $\Z (\theta )$ is continuous and non-constant, the Lyapunov exponent $\lambda$ will be negative, so the synchronous state $\theta_1 = \theta_2$ will be stable. Note, continuity is not satisfied in the case of our singular approximation to $\Z (\theta)$. We demonstrate the accuracy of our theory (\ref{lyapapprox}) in Fig. \ref{figstochsynch}{\bf C},{\bf D}, showing that $\lambda$ decreases as a function of $\tau$ and is non-monotonic in $I$. Thus, slow oscillations with longer periods are synchronized more quickly, relative to the number of oscillation cycles. Since the Lyapunov exponent has highest amplitude $|\lambda|$ for both low and high values of the tonic input $I$, we also suspect this is related to the period of the oscillation $T$.

\section{Impact of intrinsic noise on stochastic synchronization}
\label{indepnos}

\begin{figure}
\begin{center} \includegraphics[width=5.5cm]{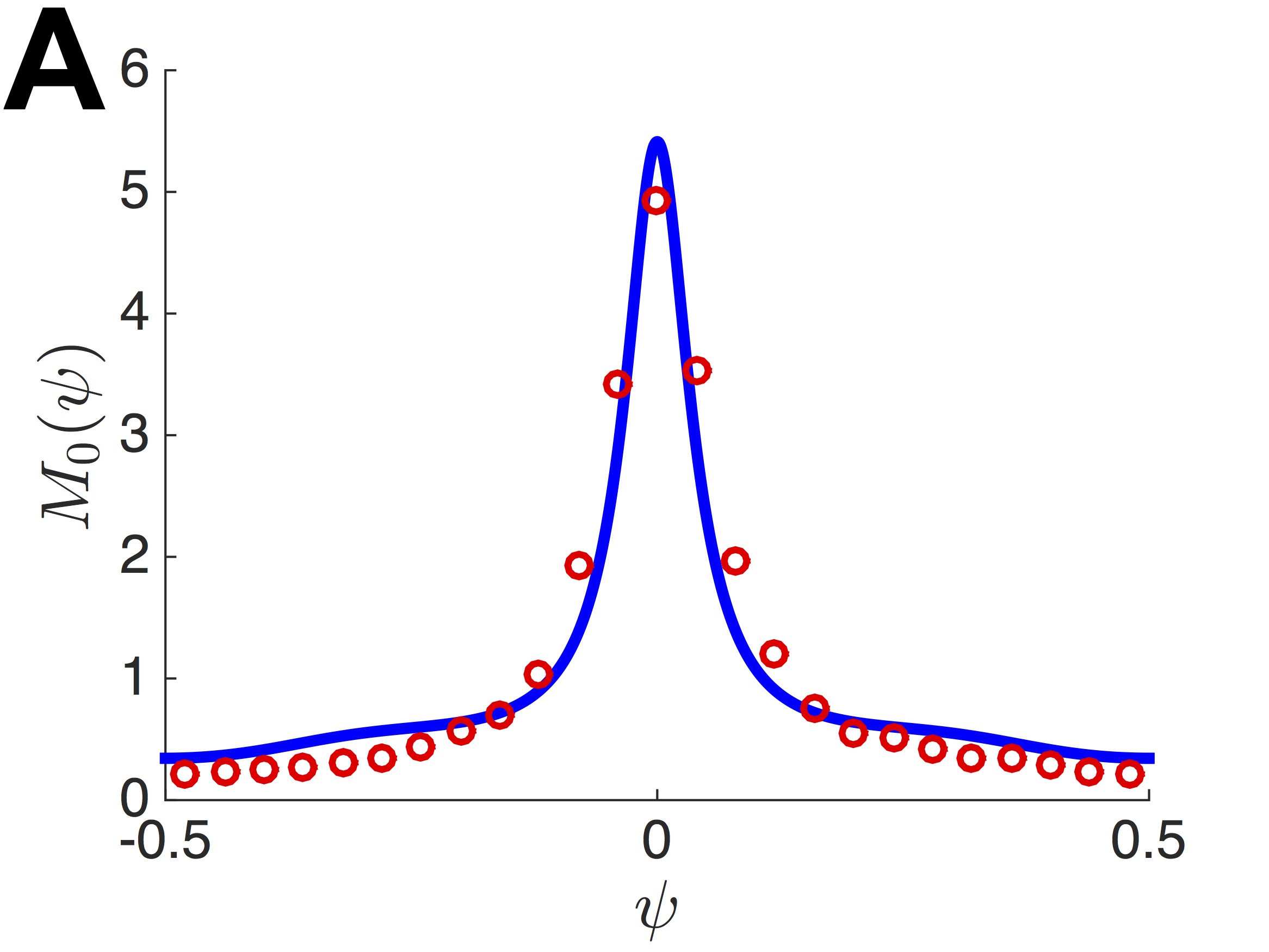} \includegraphics[width=5.5cm]{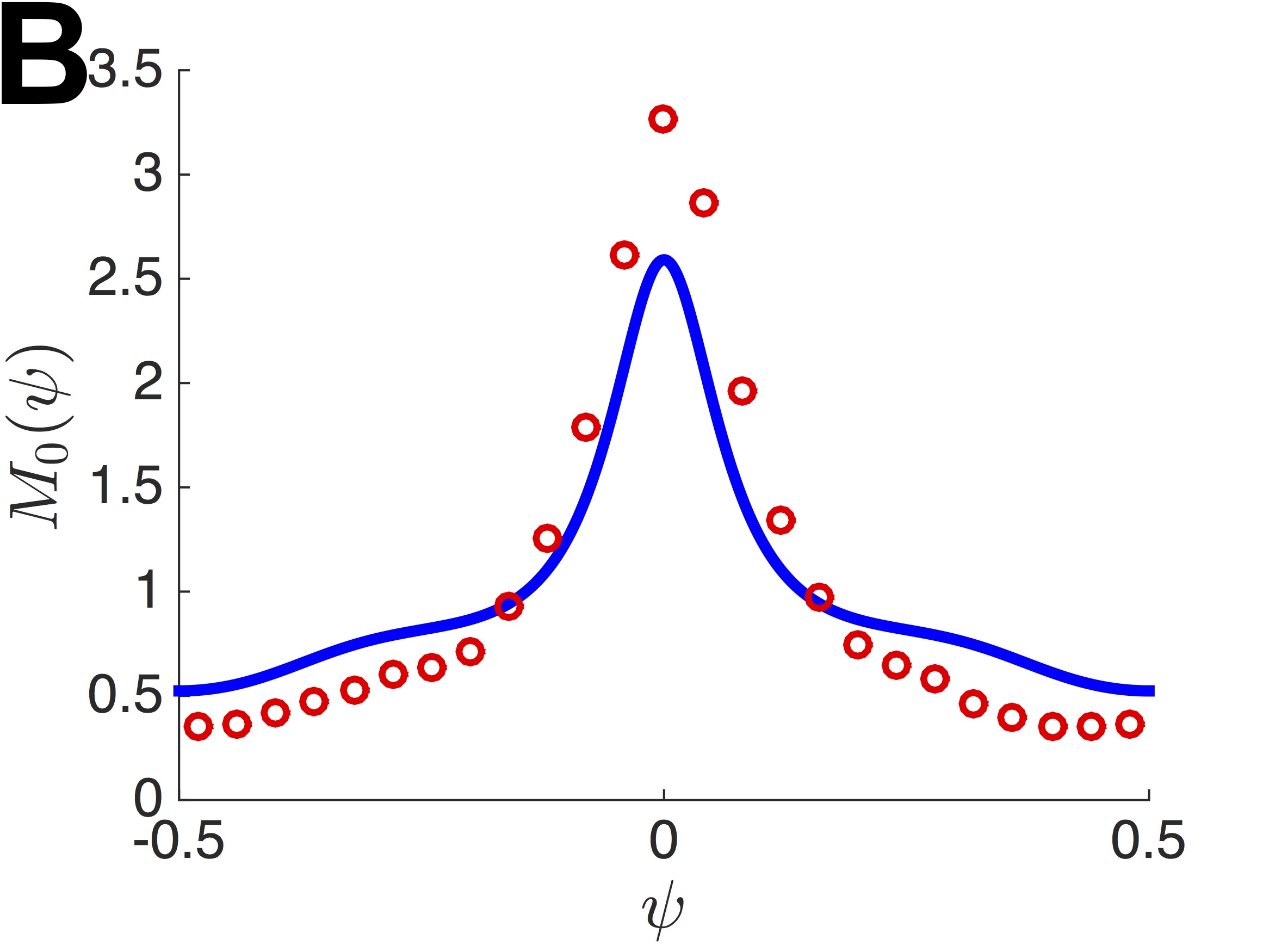}\end{center}
\caption{Stationary density $M_0(\psi)$ of the phase difference $\psi = \theta_1 - \theta_2$ for two slowly oscillating neural population driven by both common and independent noise (\ref{dualind}). As the degree of noise correlation is decreased from ({\bf A}) $\chi_a = 0.95$ to ({\bf B}) $\chi_a = 0.90$, the density spreads, but there is still a peak at $\psi = 0$, the phase-locked state. We focus on noise in the adaptation variable, so $\sigma_u = 0$ and $\sigma_a = 0.01$. Other parameters are $\alpha = 0.5$, $\gamma = 15$, $\phi = 1$, and $\tau = 20$.}
\label{figindnoise}
\end{figure}

We now extend our results from the previous section by studying the impact of independent noise in each population. Independent noise is incorporated into the modified model (\ref{dualind}). Noting, again there is a periodic solution to the noise-free version of this system, phase-reduction methods can be used to obtain approximate Langevin equations for the phase variables \citep{nakao07}
\begin{subequations} \label{indphase}
\begin{align}
\d \theta_1 &=  \omega \d t + \Z ( \theta_1(t) ) \cdot \left[  \d \bxi_c (t) + \d \bxi_1 (t) \right], \\
\d \theta_2 &= \omega \d t + \Z ( \theta_2(t) ) \cdot \left[ \d \bxi_c (t) + \d \bxi_2 (t) \right],
\end{align}
\end{subequations}
where the noise vectors $\bxi_c = (\chi_u \xi_{uc}, \chi_a \xi_{ac})^T$ and $\bxi_j = (\sqrt{1- \chi_u^2} \xi_{uj}, \sqrt{1 - \chi_a^2} \xi_{aj})^T$ ($j=1,2$). We can reformulate the pair of Stratonovich differential equations as Ito stochastic differential equations given by the system
\begin{subequations} \label{inditophase}
\begin{align}
\d \theta_1 &= A_1( \bthet ) \d t + \d \zeta_1( \bthet, t), \\
\d \theta_2 &= A_2 ( \bthet ) \d t + \d \zeta_2 ( \bthet, t),
\end{align}
\end{subequations}
where the statistics of the noise terms $\d \zeta_j( \bthet, t) =  \Z ( \theta_j(t) ) \cdot \left[  \d \bxi_c (t) + \d \bxi_j (t) \right] $ ($j=1,2$) are specified by $\langle \d \zeta_j ( \bthet, t)\rangle = 0$ and $\langle \d \zeta_j ( \bthet, t) \d \zeta_k ( \bthet, t) \rangle = C_{jk}(\bthet ) \d t$ where
\begin{align}
C_{jk} ( \bthet ) =& \left(  \chi_u \Db_{uc}^{1/2} Z_u( \theta_j ) + \chi_a \Db_{ac}^{1/2} Z_a( \theta_j) \right) \left( \chi_u \Db_{uc}^{1/2} Z_u( \theta_k) + \chi_a \Db_{ac}^{1/2} Z_a( \theta_k) \right) \nonumber \\
& + \left( \sqrt{1 - \chi_u^2} \Db_{ul}^{1/2} Z_u( \theta_j) + \sqrt{1 - \chi_a^2} \Db_{al}^{1/2} Z_a( \theta_j) \right)^2 \delta_{j,k},  \label{corrfunc}
\end{align}
separating the impact of correlated and local sources of noise. The drift terms can thus be calculated $A_j( \bthet ) = \omega + \frac{1}{4} \frac{\pd}{\pd \theta_j} C_{jj} ( \bthet )$. The Fokker-Planck equation describing the evolution of the probability density function $P( \bthet, t)$ of the phases is given
\begin{align}
\frac{\pd P}{\pd t} = - \sum_{j=1}^2 \frac{\pd}{\pd \theta_j} \left[ A_j( \bthet ) P \right] + \frac{1}{2} \sum_{j=1}^2 \sum_{k=1}^2 \frac{\pd^2}{\pd \theta_j \pd \theta_k} \left[ C_{jk} ( \bthet ) P \right].  \label{fpe1}
\end{align}
Now, we apply a change of variables to the Fokker-Planck equation (\ref{fpe1}) defined $\theta_j = \omega t + \vartheta_j$. Assuming noise is weak, the function $Q( \bvthet , t)$ varies slowly compared with the period of the phase oscillators $\theta_j$. Thus, we average the drifts $A_j( \bthet )$ and correlation function $C_{jk} ( \bthet )$ over a single period. The resulting Fokker-Planck equation is then
\begin{align*}
\frac{\pd Q( \bvthet, t)}{\pd t} = \frac{1}{2} \sum_{j=1}^2 \sum_{k=1}^2 \frac{\pd^2}{\pd \vartheta_j \pd \vartheta_k} \left[ B_{jk} ( \bvthet ) Q \right]
\end{align*}
where the averaged correlation function is given by the formula
\begin{align*}
B_{jk} ( \bvthet ) = g ( \theta_1 - \theta_2 ) + h(0) \delta_{j,k},
\end{align*}
where the correlation functions are defined
\begin{align*}
g( \theta ) = \int_0^1 \left[ \chi_u^2 \Db_{uc} Z_u( \theta') Z_u(\theta' + \theta ) + \chi_a^2 \Db_{ac} Z_a( \theta' ) Z_a( \theta' + \theta ) \right] \d \theta'
\end{align*}
and
\begin{align*}
h( \theta ) = \int_0^1 \left[ (1 - \chi_u^2) \Db_{ul} Z_u( \theta') Z_u(\theta' + \theta ) + (1 - \chi_a^2) \Db_{al} Z_a( \theta' ) Z_a( \theta' + \theta ) \right] \d \theta'.
\end{align*}
We study the relationship between the phases $\vartheta_1$ and $\vartheta_2$ by substituting the formula for the averaged correlation matrix
\begin{align}
\frac{\pd Q( \bvthet ,t)}{\pd t} = \frac{1}{2} \left[ g(0) + h(0) \right] \left[  \frac{\pd^2 Q}{\pd \vartheta_1^2} + \frac{\pd^2 Q}{\pd \vartheta_2^2} \right] + \frac{\pd^2}{\pd \vartheta_1 \pd \vartheta_2} \left[ g( \vartheta_1 - \vartheta_2) Q \right].  \label{fp2}
\end{align}
We can write (\ref{fp2}) as a separable equation by employing a change of variables that tracks the average $\rho = (\vartheta_1 + \vartheta_2)/2$ and phase difference $\psi = \vartheta_1 - \vartheta_2$ of the original position variables, so
\begin{subequations} 
\begin{align}  \label{fpsep}
\frac{\pd U ( \rho , t)}{\pd t} &= \frac{1}{4} \left[ g(0) + g( \psi ) + h(0) \right] \frac{\pd^2 U(\rho, t)}{\pd \rho^2}, \\
\frac{\pd M ( \psi ,t)}{\pd t} &= \frac{\pd^2}{\pd \psi^2} \left[ g(0) - g( \psi ) + h(0) \right] M( \psi ,t).
\end{align}
\end{subequations}
Thus, we can solve for the stationary solution of the system (\ref{fpsep}) by serving $U_t = M_t \equiv 0$ and requiring periodic boundary conditions. We find that the stationary distribution of the position average is $U_0(\rho) = 1$. In addition, we can integrate the stationary equation for $M( \psi ,t)$ to find
\begin{align} \label{M0form}
M_0( \psi ) = \frac{m_0}{\sigma_u^2 \left[ (2-\chi_u^2) g_u(0) - \chi_u^2 g_u( \psi ) \right] + \sigma_a^2 \left[ (2-\chi_a^2) g_a(0) - \chi_a^2 g_a(\psi) \right] },
\end{align}
where $m_0 = 1/ \int_0^1 \left[ g(0) - g(x) + h(0) \right]^{-1} \d x$ is a normalization factor and we have simplified the expression using $\Db_{u1} = \Db_{u2} \equiv \Db_{ul} = \sigma_u^2$ and $\Db_{a1} = \Db_{a2} \equiv \Db_{al} = \sigma_a^2$ and defined
\begin{align*}
g_j( \psi ) = \int_0^1 Z_j( \theta ) Z_j( \theta + \phi ) \d \theta.
\end{align*}
When noise to each layer is independent ($\chi_u, \chi_a \to 0$), then $M_0(\psi ) = 1$ is constant in space. When noise is totally correlated ($\chi_u, \chi_a \to 1$), then $M_0(\psi) = \delta ( \phi )$. The stationary distribution $M_0(\psi)$ will broaden as the correlations $\chi_u$ and $\chi_a$ are decreased from unity, with a peak remaining at $\phi = 0$. We demonstrate the accuracy of the formula (\ref{M0form}) for the stationary density of the phase difference $\psi$ in Fig. \ref{figindnoise}, showing that it widens as the level noise correlation is decreased. Again, we focus on the impact of adaptation noise. Thus, even when independent noise is introduced, there is some semblance of synchronization in the phases of two noise-driven neural populations (\ref{dualind}). 

\section{Discussion}
\label{disc}

We have studied the impact of deterministic and stochastic perturbations to a neural population model of slow oscillations. The model was comprised of a single recurrently coupled excitatory population with negative feedback from a slow adaptive current \citep{laing02,jayasuriya12}. By examining the phase sensitivity function $(Z_u, Z_a)$, we found that perturbations of the adaptation variable lead to much larger changes in oscillation phase than perturbations of neural activity. Furthermore, this effect becomes more pronounced as the timescale $\tau$ of adaptation is increased. Introducing noise in the model decreases the oscillation period and helps to balance the mean duration of the oscillation's up and down states. When two uncoupled populations receive common noise, their oscillation phases $\theta_1$ and $\theta_2$ eventually become synchronized, which can be shown by deriving a formula for the Lyapunov exponent of the absorbing state $\theta_1 \equiv \theta_2$ \citep{teramae04}. When independent noise is introduced to each population, in addition to common noise, the long-term state of the system is described by a probability density for $\psi = \theta_1 - \theta_2$, which peaks at $\psi \equiv 0$.

Our study was motivated by the observation that recurrent cortical networks can spontaneously generate stochastic oscillations between up and down states. Guided by previous work in spiking models \citep{compte03}, we explored a rate model of a recurrent excitatory network with slow spike frequency adaptation. One of the open questions about up and down state transitions concerns the degree to which they are generated by noise or by more deterministic mechanisms, such as slow currents or short term plasticity \citep{cossart03}. Here, we have provided some characteristic features that emerge as the level of noise responsible for transitions is increased. Similar questions have been probed in the context of models of perceptual rivalry \citep{morenobote07}. In addition, we have provided a plausible mechanism whereby the onset of up and down states could be synchronized in distinct networks \citep{volgushev06}.

There are several potential extensions to this work. For instance, we could examine the impact of long-range connections between networks to see how these interact with common and independent noise to shape the phase coherence of oscillations. Similar studies have been performed in spiking models by \cite{ly09}. Interestingly, shared noise can actually stabilize the anti-phase locked state in this case, even though it is unstable in the absence of noise. Furthermore, it is known that coupling spanning long distances can be subject to axonal delays. In spite of this, networks of distantly coupled clusters of cells can still sustain zero-lag synchronized states \citep{vicente08}. Thus, we could also explore the impact of delayed coupling, determining how features of phase sensitivity function interact with delay to promote in-phase or anti-phase synchronized states.

%\begin{acknowledgements}
%This publication was based on work supported in part by the National Science Foundation (DMS-1311755).
%\end{acknowledgements}

% BibTeX users please use one of
%\bibliographystyle{spbasic}      % basic style, author-year citations
%\bibliographystyle{spmpsci}      % mathematics and physical sciences
%\bibliographystyle{spphys}       % APS-like style for physics
\bibliographystyle{jneurosci}
\bibliography{adaptprc}

\end{document}